\documentclass{article}

\usepackage{arxiv}

\usepackage[utf8]{inputenc} 
\usepackage[T1]{fontenc}    
\usepackage{hyperref}      
\usepackage{url}            
\usepackage{booktabs}       
\usepackage{amsfonts}       
\usepackage{nicefrac}       
\usepackage{microtype} 
\usepackage{lipsum}		
\usepackage{graphicx}
\usepackage{natbib}
\usepackage{doi}
\usepackage{multirow}
\usepackage{amsmath}
\usepackage{bm}
\usepackage{appendix}
\usepackage{xcolor}

\title{Bayesian variable selection in sample selection models using spike-and-slab priors}

\author{ \href{https://orcid.org/0009-0005-0572-2381}{\includegraphics[scale=0.06]{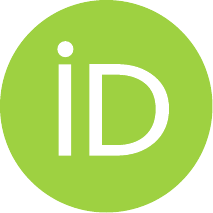}\hspace{1mm}Adam J. Iqbal}\\
	Department of Mathematical Sciences\\
	Durham University\\
	Durham, DH1 3LE \\
	\texttt{adam.iqbal@durham.ac.uk} \\
	\And
	\href{https://orcid.org/0000-0001-9252-9275}{\includegraphics[scale=0.06]{orcid.pdf}\hspace{1mm}Emmanuel O. Ogundimu} \\
	Department of Mathematical Sciences\\
	Durham University\\
	Durham, DH1 3LE\\
	\texttt{emmanuel.ogundimu@durham.ac.uk} \\
	\And
	\href{https://orcid.org/0000-0001-7183-8407}{\includegraphics[scale=0.06]{orcid.pdf}\hspace{1mm}F. Javier Rubio} \\
	Department of Statistical Science\\
	University College London\\
	London, WC1E 7HB\\
	\texttt{f.j.rubio@ucl.ac.uk} \\
}

\renewcommand{\shorttitle}{Bayesian variable selection in sample selection models using spike-and-slab priors}

\hypersetup{
pdftitle={Bayesian variable selection in sample selection models using spike-and-slab priors},
pdfsubject={stat.CO, stat.ME},
pdfauthor={Adam Iqbal, Emmanuel Ogundimu, F. Javier Rubio},
pdfkeywords={Gibbs sampling, missing data, sample selection, spike-and-slab prior},
}

\begin{document}
\maketitle

\begin{abstract}
Sample selection models are a widely used approach for correcting bias caused by data that are missing not at random. Their formulation requires specifying the variables that influence the outcome and those that drive the selection process. This specification is often based on expert knowledge, which can result in the inclusion of irrelevant variables or the omission of important ones. Moreover, to avoid inferential problems such as practical non-identifiability, practitioners frequently impose \emph{exclusion restrictions}, that is, model specifications in which certain variables predict selection but have no effect on the outcome of interest.
A recent proposal employs adaptive LASSO to select the variables that enter into the outcome and selection equations, but its performance depends on the so-called covariance assumption, which can be violated in small to moderate samples. To address these challenges, we propose two families of spike-and-slab priors to conduct Bayesian variable selection in sample selection models. These prior structures allow for constructing a Gibbs sampler with tractable conditionals, which is scalable to the dimensions of practical interest.
We illustrate the performance of the proposed methodology through a simulation study and present a comparison against adaptive LASSO and stepwise selection. We also provide two applications using publicly available real data.
\end{abstract}



\section{Introduction}\label{sec1}

Sample selection occurs when a portion of the sample is missing non-randomly, leading to a sample that is not representative of the population of interest. This is prevalent in medical and social sciences, where the outcome of interest is non-randomly selected \citep{heckman1976common,heckman1979sample,wooldridge2010econometric}. As a result, the sample becomes biased, potentially distorting the findings and generalizability of the results. 
\cite{heckman1976common,heckman1979sample} introduced a regression model to correct for sample selection bias in the case where the outcome variable is continuous. The main idea is to formulate a two-equation regression model, one for the outcome process and one for the selection process, allowing for correlated errors under the assumption of bivariate normality. These kind of models are known as \textit{sample selection} models (or Heckman correction). \cite{heckman1979sample} proposed a two-stage estimator of the parameters of sample selection models and established the consistency and asymptotic normality of such an estimator. After this influential development, several extensions have been proposed to alleviate challenges posed by the assumption of normality of the errors. These include
\cite{van2011bayesian}, who extended the outcome model to mixture-of-Gaussian error distributions.
\cite{marchenko2012heckman}, who extended Heckman's model by modelling the errors in the two-equation model using a bivariate-$t$ distribution, which accounts for heavy tails, coupled with maximum likelihood estimation of the parameters. \cite{ogundimu2016sample} assumed bivariate skew-normal errors, which account for departures from symmetry. Other extensions include the analysis of binary outcomes subject to sample selection bias.
One of the challenges in these models is that the user must specify the variables that enter the outcome and selection models. Entering the same variables into both the outcome and selection submodels often leads to collinearity problems. 
It has also been shown that even with ``exclusion restriction'' rules - a covariate included in the selection but not the outcome - collinearity issues may  appear in two-step estimators \citep{leung2000collinearity}. 
This challenge points to the need for formal tools for specifying the variables that enter the outcome and selection models.


Recently, the use of variable selection methods in the context of sample selection models has attracted considerable attention. For instance, \cite{ogundimu2022regularization} used an Adaptive LASSO penalty to select the variables that enter the outcome and selection models. 
The distributional regression sample selection modelling introduced by \cite{wiemann2022correcting} allows for the use of Bayesian variable selection priors (shrinkage and spike-and-slab priors). However, \cite{wiemann2022correcting} only provide an implementation of their proposed model using a Metropolis within Gibbs sampler, which is computationally more expensive than a closed-form Gibbs sampler due to the need for iterative accept-reject Metropolis steps, and requires careful calibration of the priors for the additive components to ensure convergence.
In a similar vein, \cite{vera2023bayesian} considered the use of shrinkage and spike-and-slab priors, but they restricted variable selection to only the selection equation, and their implementation similarly relies on a Metropolis within Gibbs sampler.


We propose two families of continuous spike-and-slab priors \citep{george1993variable,tadesse2021handbook} to conduct Bayesian variable selection in sample selection models. The first class of priors allows for the use of scale mixtures of normals for the spike and slab components, which include Laplace and Student-$t$ distributions. The second one is defined conditionally on the marginal variance of the outcome equation, allowing for easier calibration, and also includes scale mixtures of normals for the spike and slab components. We provide a detailed discussion on the calibration of both priors, taking into account the different nature of the outcome and selection models. 
We derive a closed-form Gibbs sampler for the posterior distribution of the parameters, under both classes of priors, which scales well to the dimensions of interest in practical applications.
Posterior samples can be used to jointly select variables for both the selection and outcome equations, without the need for an exclusion restriction. 

The remainder of this paper is organized as follows. In Section \ref{sec:sampsel}, we present a brief review of sample selection models and previous literature on Bayesian inference for these models. In Section \ref{sec:priors}, we present the proposed families of spike-and-slab priors and discuss the calibration of these priors.
Section \ref{sec:gibbs} presents details about the Gibbs sampler for the proposed prior structures. Section \ref{sec:simulation} presents a simulation study aimed at illustrating the performance of the proposed spike-and-slab priors for variable selection. We compare the performance of the proposed methodology with that of adaptive LASSO and forward selection. Indeed, a byproduct of this work is the implementation of stepwise variable selection methods for sample selection models.
Section \ref{sec:applications} presents real data applications using two popular data sets in the context of sample selection models and a sensitivity analysis for the hyperparameters. We conclude with a discussion and possible extensions in Section \ref{sec:discussion}. An implementation and code to reproduce the results in this paper can be found at \url{https://github.com/adam-iqbal/selection-spike-slab}

\section{Sample selection models}\label{sec:sampsel}

{Sample selection models address non-random missing outcomes by coupling a latent outcome equation with a latent selection index. Let $s_i\in\{0,1\}$ denote whether $y_i^{\ast}$ is observed ($s_i=1$) or missing ($s_i=0$), and let $s_i^{\ast}$ be the unobserved propensity for selection (\textit{e.g.}, an individual's propensity to participate in the labor force).} The model comprises
\begin{eqnarray}
y^{*}_{i} &=& \beta_{0} + \bm{x}_{i}^{\top}\bm{\beta} + \epsilon_{1,i} \, , \quad \text{(outcome)} \nonumber\\
s^{*}_{i} &=& \alpha_{0} + \bm{w}_{i}^{\top}\bm{\alpha} + \epsilon_{2,i}\, , \quad \text{(selection)}
\label{eq:twoeqsampsel}
\end{eqnarray}

with observed data $(y_i,s_i)$ related by $s_{i}=\mathbb{I}(s^{\ast}_{i}>0)$ and
\begin{eqnarray}\label{eq:responses}
    y_{i}=
\begin{cases}
y_{i}^{\ast}, & s_{i}=1,\\
\text{not observed}, & s_{i}=0.
\end{cases}
\end{eqnarray}

Here $\bm{x}_{i}=(x_{i,1},\ldots,x_{i,p})^{\top}\in\mathbb{R}^{p}$ and $\bm{w}_{i}=(w_{i,1},\ldots,w_{i,q})^{\top}\in\mathbb{R}^{q}$, $i=1,\ldots,n$, are covariates; $\bm{\beta}=(\beta_{1},\ldots,\beta_{p})^{\top}\in\mathbb{R}^{p}$ and $\bm{\alpha}=(\alpha_{1},\ldots,\alpha_{q})^{\top}\in\mathbb{R}^{q}$ are regression coefficients; and $\beta_{0},\alpha_{0}$ are intercepts. In the classical sample selection model, the bivariate errors are assumed to be independent and identically distributed, following a bivariate normal distribution:
\begin{eqnarray}
\begin{pmatrix}
\varepsilon_{1,i} \\
\varepsilon_{2,i}
\end{pmatrix}
\sim \mathcal{N}\!\left(
\begin{pmatrix}
0\\
0
\end{pmatrix},
\begin{pmatrix}
\sigma^{2} & \sigma \rho \\
\sigma \rho & 1
\end{pmatrix} \right), \label{corr_matrix}
\end{eqnarray}
with $\sigma>0$ and $\rho\in(-1,1)$. In \eqref{corr_matrix} the selection variance is fixed to $1$ because only the sign of $s_i^{\ast}$ is observed, so the index scale is not identified (\textit{e.g.}, \citealp[Ch.~17]{Greene2018}; \citealp[Ch.~15]{wooldridge2010econometric}). The selection regression $s_i^{\ast}=\alpha_{0}+\bm{w}_{i}^{\top}\bm{\alpha}+\varepsilon_{2,i}$ implies $\Pr(s_i=1\mid\bm w_i)=\Phi(\alpha_{0}+\bm{w}_{i}^{\top}\bm{\alpha})$ (a probit model for non-missingness).

If $\rho=0$, selection is non-informative and the data is missing at random (MAR) given the covariates, and valid inference for the conditional distribution
of $y^{\ast}$ given $\bm x$ can be based on the complete cases (or by adjusting for covariates). In this case the outcome error has zero mean in the selected sample, and $\mathbb{E}(\varepsilon_{1,i}\mid s_i=1)=0$.  Thus, a standard regression on the observed data is consistent. If, in addition, the non-intercept terms in the selection equation coefficients (\textit{i.e.}, $\bm\alpha$) are zero, then selection depends only on an intercept, and in this case the data is missing completely at random (MCAR), and no adjustment is needed.

If $\rho>0$, we have positive selection bias and the implication is that unobserved factors that make selection more likely are positively correlated with unobserved factors that increase the outcome. So the selected sample tends to have a higher mean outcome than a random draw from the population. Conversely, there is negative selection when $\rho<0$. In general, when $\rho\neq 0$, the outcome error has a nonzero conditional mean among selected units, which is the source of the bias. Using the properties of the bivariate normal distribution, this conditional mean can be derived as:
\begin{equation*}
\mathbb{E}(\varepsilon_{1,i}\mid s_i=1,\bm x_i,\bm w_i)
= \mathbb{E}\!\left(\varepsilon_{1,i}\mid \varepsilon_{2,i} > -(\alpha_0+\bm w_i^{\top}\bm\alpha)\right)
= \sigma\,\rho\,\lambda\!\left(\alpha_0+\bm w_i^{\top}\bm\alpha\right),
\end{equation*}
where $\lambda(u)=\phi(u)/\Phi(u)$ is the inverse Mills ratio (IMR) and $\phi,\Phi$ denote the standard normal probability density function (pdf) and cumulative distribution function (cdf). The IMR equals the mean of a standard normal truncated to $\{Z>-u\}$; it is strictly positive and strictly decreasing in $u$ (indeed, $\lambda'(u)=-\lambda(u)\{u+\lambda(u)\}<0$). Larger selection propensity $u=\alpha_0+\bm w_i^\top\bm\alpha$ therefore implies a smaller correction term $\sigma\rho\,\lambda(u)$. Consequently,
\begin{eqnarray}\label{TS}
\mathbb{E}(y_i^{\ast}\mid s_i=1,\bm x_i,\bm w_i)
=\beta_0+\bm x_i^{\top}\bm\beta+\sigma\,\rho\,\lambda\!\left(\alpha_0+\bm w_i^{\top}\bm\alpha\right).
\end{eqnarray}

Beyond the mean, $\rho$ also affects dispersion in the selected sample. For the truncated bivariate normal one obtains $\mathrm{Var}(\varepsilon_{1,i}\mid s_i=1,\bm x_i,\bm w_i) =\sigma^2\!\left[1-\rho^2\,\kappa(u_i)\right]$, $u_i:=\alpha_0+\bm w_i^\top\bm\alpha$, and $\kappa(u):=\lambda(u)\{\lambda(u)+u\}$.
This implies that the selection process makes the observed data heteroscedastic, and the observed variance is always less than the true population variance $\sigma^2$. The amount of this variance reduction depends on both the strength of the selection effect ($\rho$) and each observation's propensity to be selected ($u_i$).

Equation \eqref{TS} is the conditional expectation of the observed data and is the basis of the Heckman two-step method \citep{heckman1979sample}. In practice, a probit for $s_i$ on $\bm w_i$ provides estimates $(\hat\alpha_0,\hat{\bm\alpha})$, from which $\widehat{\lambda}_i=\lambda(\hat\alpha_0+\bm w_i^{\top}\hat{\bm\alpha})$ is computed. One then estimates, on the selected sample $\{i:s_i=1\}$,
\[
y_i \;=\; \beta_0+\bm x_i^{\top}\bm\beta \;+\; \delta\,\widehat{\lambda}_i \;+\; \text{error},
\]
where the coefficient $\hat\delta$ estimates $\sigma\rho$.

The log-likelihood of the parameters for the sample selection model \eqref{eq:twoeqsampsel}--\eqref{corr_matrix} is given by \citep{toomet:2008}
\begin{eqnarray*}
\ell(\alpha_{0}, \beta_{0}, \bm{\alpha}, \bm{\beta}, \sigma, \rho) &=& \sum_{\{s_{i}=0\}}\log\Phi(-\alpha_{0} - \bm{w}_{i}^{\top}\bm{\alpha})\\
&+& \sum_{\{s_i=1\}} \Bigg( \log\Phi\left( \frac{\alpha_{0} + \bm{w}_{i}^{\top}\bm{\alpha} + \rho(y_{i} - \beta_{0} - \bm{x}_{i}^{\top}\bm{\beta})/\sigma}{\sqrt{1-\rho^{2}}}\right)\\ && \qquad \quad - \frac{1}{2}\left(\frac{y_{i} - \beta_{0} - \bm{x}_{i}^{\top}\bm{\beta}}{\sigma}\right)^{2} - \log(\sigma) - \frac{1}{2}\log(2\pi) \Bigg) \, .
\end{eqnarray*}

Maximum likelihood estimators of the parameters can be found by maximizing the log-likelihood function using any general-purpose optimization method.

A central challenge is specifying covariates for the outcome ($\bm x_i$) and selection ($\bm w_i$) equations. In practice, those variables that affect the outcome also affect the selection, so $\bm x_i$ is often chosen either as a subset of, or equal to, $\bm w_i$. While the model is identified in principle even if $\bm x_i$ and $\bm w_i$ coincide, practical estimation is often hindered by flat likelihood regions and potential misspecification. Empirical identification is therefore substantially stronger when an \emph{exclusion restriction} (ER) is available (\textit{e.g.}, \citealp{chib2009estimation,wiesenfarth2010bayesian,vella1998estimating}). In many applications - with samples typically in the hundreds to low thousands (\textit{e.g.}, Mroz, $n{=}753$, Ambulatory expenditures data, $n{=}3,328$) - the same predictors plausibly influence both selection and outcome. This overlap, coupled with the fact that the inverse Mills ratio $\lambda(u)$ is nearly linear over wide ranges of its support, can induce severe multicollinearity, particularly in two-step implementations \citep{puhani:2000, leung2000collinearity}.

When theory points to identical covariates, a common reaction is a ``mad'' search for an ER, but adding extraneous regressors solely for identification risks specification error \citep{sartori2003estimator}. We therefore propose a joint, data-adaptive variable-selection approach across both equations. This framework allows plausible exclusions to be formally tested rather than imposed, while shrinkage helps mitigate collinearity when $\bm x_i$ and $\bm w_i$ necessarily overlap.

In the Bayesian framework, \cite{van2011bayesian} showed that it is possible to construct a tractable Gibbs sampler for the model defined by equations \eqref{eq:twoeqsampsel}-\eqref{eq:responses} by using the reparametrization $\tilde{\sigma}^{2} = \sigma^{2}(1 - \rho^{2})$ and $\tilde{\rho} = \rho\sigma$.
The conditional distributions for $(s^{*}_{i}, y^{*}_{i})$ under this reparametrization are
\begin{equation}
\begin{split}
s^{*}_{i} \mid \bm{\theta}  &\sim
N(\alpha_{0} + \bm{w}^{\top}_{i}\bm{\alpha}, 1) \, ,\\
s^{*}_{i} \mid \{y^{*}_{i}, \bm{\theta}\}  &\sim
N(\alpha_{0} + \bm{w}^{\top}_{i}\bm{\alpha} + \frac{\tilde{\rho}}{\tilde{\sigma}^{2} + \tilde{\rho}^{2}}(y_{i} - \beta_{0} - \bm{x}^{\top}_{i}\bm{\beta}), \frac{\tilde{\sigma}^{2}}{\tilde{\sigma}^{2} + \tilde{\rho}^{2}})\, ,\\
y^{*}_{i} \mid \{s^{*}_{i}, \bm{\theta}\} &\sim N\left(\beta_{0} + \bm{x}^{\top}_{i}\bm{\beta} + \tilde{\rho}(s^{*}_{i} - \alpha_{0} - \bm{w}_{i}^{\top}\bm{\alpha}\right), \tilde{\sigma}^{2})\, . \label{likelihoods}
\end{split}
\end{equation}
where $\bm{\theta} = (\beta_{0},\bm{\beta}^{\top},\alpha_{0},\bm{\alpha}^{\top},\tilde{\rho},\tilde{\sigma}^{2})^{\top}$.
Full details about the Gibbs sampler, and the corresponding priors, are presented by \cite{van2011bayesian} and reproduced in Appendix \ref{sec:vansampler} for completeness.
The major advantage of this Gibbs sampler is that every conditional distribution is in closed-form, allowing for fast simulations. The sampler does rely on $(p+1) \times (p+1)$ and $(q+1) \times (q+1)$ matrix inversions, but it is rare for the problems of interest in sample selection literature to exhibit dimensions high enough for this to pose a serious concern. The Gibbs sampler also requires sampling from $n$ truncated normal random variables at each step, but this can be done relatively fast using available methods and R packages \citep{geweke1991efficient}. We will capitalize on \cite{van2011bayesian} to develop a Gibbs sampler in Section \ref{sec:gibbs} for the priors proposed in Section \ref{sec:priors}.

\section{Spike-and-slab prior formulation}\label{sec:priors}
Our aim is to perform Bayesian variable selection in sample selection models of type \eqref{eq:twoeqsampsel}. To this end, let us define the variable inclusion indicators as follows. For $j = 1, \ldots, p$, let $\gamma^{O}_{j} = 1$ if $\beta_{j}$ is included in the outcome model, and $\gamma^{O}_{j} =0$ otherwise. Similarly, for $k = 1, \ldots, q$, let $\gamma^{S}_{k} = 1$ if $\alpha_{k}$ is included in the selection model, and $\gamma^{S}_{k} = 0$ otherwise. Next, we present the two proposed classes of priors. The first class of priors contains continuous spike-and-slab priors with scale mixture of normal components \citep{george1993variable}. The second class of priors is defined conditionally on the marginal variance of the outcome model. Finally, we present a discussion on elicitation for each class of priors.



\subsection{Priors}
The Class I spike-and-slab prior \citep{george1993variable} is defined by the structure:
\begin{equation}
\begin{split}
 \bm{\beta} \mid \{\bm{\gamma^{O}}, \bm{v^{O}}\} &\sim \prod_{j=1}^{p}\left((1-\gamma^{O}_{j})N(0,\tau_{0,\beta}^{2}v^{O}_{j}) + \gamma_{j}^{O}N(0,\tau_{1,\beta}^{2}v^{O}_{j})\right) \,\\
\bm{\alpha} \mid \{\bm{\gamma^{S}},\bm{v^{S}}\} &\sim \prod_{k=1}^{q}\left((1-\gamma^{S}_{k})N(0,\tau_{0,\alpha}^{2}v^{S}_{k}) + \gamma_{k}^{S}N(0,\tau_{1,\alpha}^{2}v^{S}_{k})\right) \, ,
\end{split} \label{priors}
\end{equation}
where $v_{j}^{O}$ and $v_{k}^{S}$ are positive random variables, allowing for scale mixtures of normal distributions as priors, such as Laplace and Student-$t$ distributions which have both seen use for spike-and-slab priors in other contexts \citep{rovckova2018spike, scheipl2012spike}. Formally, we let
\begin{equation}
p(v_{j}^{O} \mid \gamma_{j}^{O}) = (1-\gamma_{j}^{O})\pi_{0,\beta}(v_{j}^{O}) + \gamma_{j}^{O}\pi_{1,\beta}(v_{j}^{O}) \, , \label{eq:mixvar}   
\end{equation} where $\pi_{0,\beta}$ and $\pi_{1,\beta}$ are probability density functions with positive support. The distribution of $v_{k}^{S} \mid \gamma_{k}^{S}$ is defined analogously.
Note that conditional on $\gamma_{j}^{O}$, only one of the components in \eqref{eq:mixvar} is active, so marginalizing \eqref{priors} over $\bm{v^{O}}$ and $\bm{v^{S}}$ results in the components being different scale mixtures of normals, depending on the choices of $\pi_{0,\beta}, \pi_{1,\beta}, \pi_{0,\alpha}$ and $\pi_{1,\alpha}$. Hence, this formulation allows for different distributions for the spike and slab components. For instance, one may desire to use a heavier-tailed distribution for the slab component than the spike component, such as using a Laplace slab with a normal spike, or alternatively using a prior with more mass concentrated around zero for the spike. As shown in Section \ref{sec:gibbs}, the only additional complexity this adds to the sampling procedure is an extra step sampling $v_{j}^{O} \mid (\gamma_{j}^{O}, \beta_{j})$ and $v_{k}^{S} \mid (\gamma_{k}^{S}, \alpha_{k})$. If these conditional distributions are closed form (as they are for Laplace and t-distribution priors) or easy to sample from, scale mixture priors can be seamlessly integrated into the sampling algorithm with minimal adverse effect on computational time.

The Class II spike-and-slab prior is defined by the structure:
\begin{equation}
\begin{split}
 \bm{\beta} \mid \{\bm{\gamma^{O}}, \bm{v^{O}}, \tilde{\sigma}^{2}\} &\sim \prod_{j=1}^{p}\left((1 - \gamma^{O}_{j})N(0,\tau_{0,\beta}^{2}v^{O}_{j}\tilde{\sigma}^{2}) + \gamma_{j}^{O} N(0,\tau_{1,\beta}^{2}v^{O}_{j}\tilde{\sigma}^{2})\right) \, ,\\
\bm{\alpha} \mid \{\bm{\gamma^{S}},\bm{v^{S}}\} &\sim \prod_{k=1}^{q}\left((1-\gamma^{S}_{k})N(0,\tau_{0,\alpha}^{2}v^{S}_{k}) + \gamma_{k}^{S}N(0,\tau_{1,\alpha}^{2}v^{S}_{k})\right) \, ,
\end{split} \label{sigma.tilde-priors}
\end{equation}
with $\bm{v^{O}}$ and $\bm{v^{S}}$ defined as before. 
In sample selection models, the variance of the observed outcomes is $\tilde{\sigma}^{2}$, as shown in equation \eqref{likelihoods}. Choosing the marginal variance $\sigma^{2} = \tilde{\sigma}^{2} + \tilde{\rho}^{2}$ would not lead to a natural expression for the posterior of $\tilde{\sigma}^{2}$, whereas $\tilde{\sigma}^{2} = \sigma^{2}(1 - \rho^{2})$ does.

The advantage of Class II priors over Class I is that the magnitude of variables is only considered relative to the unexplained variance. This is a common strategy in the standard linear regression context \citep{louzada:2023}. The cost of using $\tilde{\sigma}^{2}$ instead of $\sigma^{2}$ is that the observed variance is less than the true variance, with $\tilde{\sigma}^{2}$ being shrunk dependent on $\rho$. When $\rho = 0$, $\tilde{\sigma}^{2}$ collapses to  $\sigma^{2}$ as expected. For larger magnitudes of $\rho$, the difference is greater. For instance, when $\rho = 0.7$, $\tilde{\sigma}^{2} = 0.49\sigma^{2}$, shrinking the slab variance by a factor of about a half, which could lead to higher false positive rates.

The priors for the remaining parameters, common to both classes of priors, are specified as follows. We let $\gamma^{O}_{j}$ and $\gamma^{S}_{k}$ follow \textit{i.i.d.}~Bernoulli distributions with parameter $r$. We let $r \sim Beta(a_{0},b_{0})$, so that the priors on $\bm{\gamma}^{O}$ and $\bm{\gamma}^{S}$ are Beta-Binomial. 
One could also consider different $r$'s for the outcome and selection models, aiming at separating the penalties on the corresponding model sizes. We do not pursue this option in this work, but this could be explored as a future extension of this work.
We further assume $\tilde{\rho} \mid \tilde{\sigma} \sim N(0, \tau \tilde{\sigma}^{2})$, and $\tilde{\sigma}^{2} \sim \text{IG}(c,d)$, where $\text{IG}$ denotes an inverse gamma distribution. For the intercepts, we use weakly informative priors $\beta_{0} \sim N(0,\eta^{O}v^{O}_{0})$ and $\alpha_{0} \sim N(0,\eta^{S}v^{S}_{0})$, with large values of $\eta^{S}$ and $\eta^{O}$, and $v^{O}_{0} \sim \pi_{I,\beta}$ and $v^{S}_{0} \sim \pi_{I,\alpha}$ to allow for scale mixtures of normals as priors for the intercepts. 

\subsection{Prior calibration}\label{subsec:calibration}
Posterior inference is sensitive to the choice of both the Beta-Binomial parameters and the spike and slab variances \citep{tadesse2021handbook}. In this section we discuss intuitive choices that we found to work reasonably well in practice. The performance of this strategy will be illustrated through a simulation study in Section \ref{sec:simulation} along with a sensitivity analysis in Section \ref{sec:sens}.

\begin{figure}[h]
\centering
\includegraphics[width=0.7\textwidth]{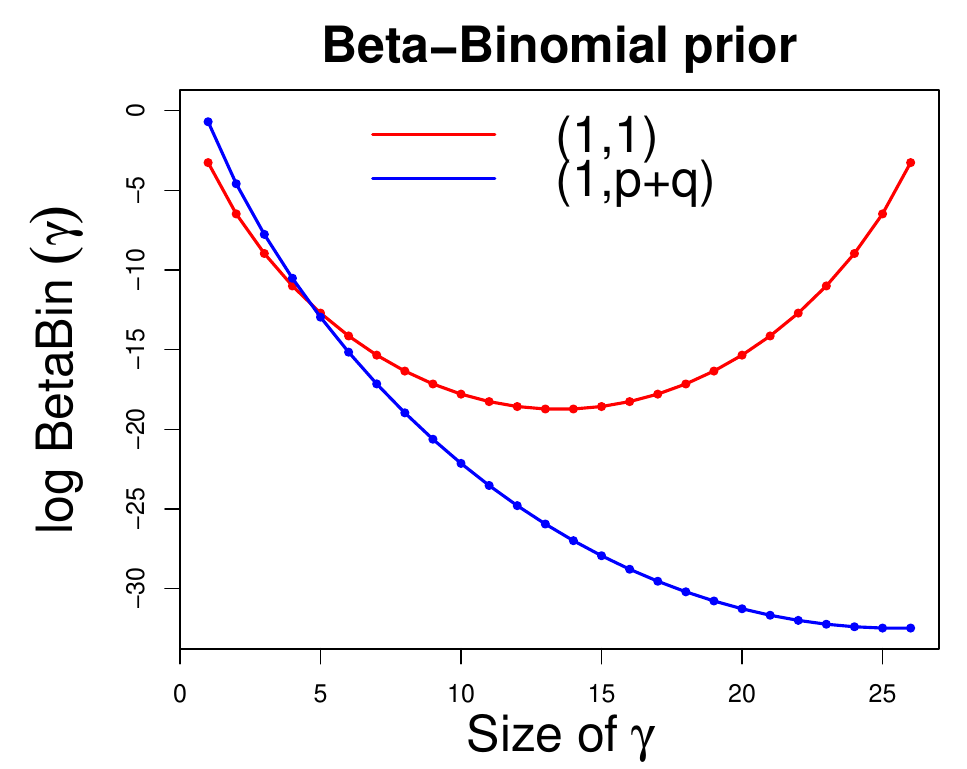}
\caption{Beta-Binomial log-prior for full model size $25$ under two different calibrations. The $x$-axis shows model size while the $y$-axis shows the prior probability of an individual model of that size.}\label{fig:betabin}
\end{figure}

\textbf{Beta-Binomial hyperparameters:} For the Beta-Binomial part of the model, $r \sim Beta(a_{0}, b_{0})$, a natural choice is $a_{0} = b_{0} = 1$, which induces a uniform prior on model size. 
This choice is reasonable in sparse scenarios, but it allows for $\mathbb{E}(r \mid \bm{\gamma^{O}},\bm{\gamma^{S}}) > 0.5$, that is the prior inclusion probability to be greater than $0.5$. In Figure \ref{fig:betabin}, which displays the Beta–Binomial prior over model sizes, the prior mass decreases as the model size increases, but begins to rise again for sizes greater than approximately $\frac{p+q}{2}$. This increase reflects the multiplicity of larger models, which receive greater individual prior probability due to the smaller number of possible configurations (analogous to models with small number of variables). Another interpretation can be found by looking at the conditional posterior of the Bernoulli parameter $r$ in Section \ref{sec:gibbs}. Under a Beta prior with parameters $(1,1)$, if more than half of the variables are included, the expected value of $r$ for the next draw exceeds $0.5$, inducing a bias towards including irrelevant variables.

An alternative is using $a_{0} = 1, b_{0} = p+q$, so that $\mathbb{E}(r \mid \bm{\gamma^{O}},\bm{\gamma^{S}}) \leq 0.5$ in every case except the full model. This can also be seen in Figure \ref{fig:betabin}: the prior includes only the left half of the parabola, so that it is monotonically decreasing. The trade-off is that significantly more sparsity is induced. As such, the choice of Beta-Binomial prior parameters depends on the context. In cases where significant sparsity is present, the choice $a_{0}=b_{0}=1$ works well and has a greater chance of including small effects. If close to (or more than) half the variables may be significant, the prior $a_{0}=1,b_{0}=p+q$ is a more sensible choice, at the cost of potentially shrinking small effects. Regardless of our recommendations, our implementation allows the user to choose their desired specification of Beta-Binomial hyperparameters.

\textbf{Spike-and-slab hyperparameters:} Continuous spike-and-slab priors are sensitive to the choice of variances, so we give particular importance to their calibration. We calibrate the variances for the normal prior case, as calibration for scale mixtures of normals will depend on the scale distribution and their tail-weight.
While theoretical results are not available for sample selection models, we take inspiration from \cite{narisetty2014bayesian} and impose $\tau_{0,\alpha} \rightarrow 0$ as $n \rightarrow \infty$ (and similarly for $\tau_{0,\beta}$). 
In sparse situations, the Beta-Binomial prior imposes greater bias against small effects as $p$ and $q$ increase. This is because each $\gamma^{O}_{j}$ and $\gamma^{S}_{k}$ are drawn from a Bernoulli with the parameter $r$ having a Beta hyperprior over it. The posterior for $r$ depends on the number of variables included - the less variables included, the smaller the expected value of $r \mid \{\bm{\gamma^{O}}, \bm{\gamma^{S}}\}$. For very sparse situations, $r$ is shrunk to small values, so that the prior inclusion probability becomes very small. As a result, the Beta-Binomial prior avoids the multiple testing problem when there are many irrelevant variables, but for small sample sizes it can lead to the exclusion of small effects more often than desired.
To partially offset this, we propose the choice $\tau_{0,\alpha} = 1/\sqrt{nc_{p}}$ and $\tau_{0,\beta} = 1/\sqrt{nc_{q}}$, with $c_{p}$ and $c_{q}$ increasing with respect to $p$ and $q$. This puts less density away from $0$ as the dimension increases, and empirically we found that $c_{p} = p$ and $c_{q} = q$ were relatively effective as heuristic choices. The factor $1/\sqrt{n}$ is motivated by theoretical results for linear regression models for spike-and-slab priors \citep{narisetty2014bayesian,narisetty:2019}. 

We similarly choose $\tau_{1,\alpha}$ and $\tau_{1,\beta}$ in line with previous work on spike-and-slab priors. For the probit equation, we choose $\tau_{1,\alpha} = C_{\alpha}$, a constant. This is in line with the choice made by \cite{narisetty:2019} for logistic regression. We choose $C_{\alpha} = \sqrt{3}/\pi$ to match the probit scale.
For the linear part of the model, we use $\tau_{1,\beta} = C_{\beta}(\log(n))^{1/2}$, which was used by \cite{narisetty2014bayesian} in their simulations. The discussion in Section 3.4 of \cite{tadesse2021handbook}  suggests that for this choice, as $n \rightarrow \infty$, the probability of including a non-zero effect will converge to $1$. This would be true for a fixed choice $\tau_{1,\beta}$ independent of $n$ as well, but the elicitation of $\tau_{1,\beta}$ is dependent on sample size. Even for moderately large sample sizes, a small choice of $\tau_{1,\beta}$ can shrink large coefficients and negatively affect the inference. But for small sample sizes, large $\tau_{1,\beta}$ will exclude small effects, losing sensitivity. Since this is not as much of an issue for moderately large sample sizes, we suggest a slab variance proportional to $\log(n)$.

The elicitation of $C_{\beta}$ still depends on the context, and our default choice is derived as follows. Suppose that $\tau_{0,\beta} = 1/\sqrt{n}$ and assume that $r = 0.5$. For a small effect of magnitude $0.25$, there is variability in the posterior sampling, so we wish to ensure that we include marginally smaller effects, say of magnitude $0.15$, with high probability, but not effects significantly smaller. With $\tau_{1,\beta} = 0.5$, the probability of including a coefficient sample of magnitude $0.15$ is approximately $96\%$. But an effect size of $0.05$ only has a $14\%$ probability of inclusion. As a result, we recommend the heuristic choice of $C_{\beta} = \left(4\log(500)\right)^{-1/2}$.

\begin{table}[ht]
\centering
\begin{tabular}{|c|cc|cc|c|}
\hline
Parameter & $\tau_{0,\beta}$ & $\tau_{0,\alpha}$ & $\tau_{1,\beta}$ & $\tau_{1,\alpha}$ & Beta-Binomial\\
\hline
Elicitation & $(np)^{-0.5}$ & $(nq)^{-0.5}$ & $0.5(\log(n)/\log(500))^{-1/2}$ & $\sqrt{3}/\pi$ & Either $(1,1)$ or $(1,p+q)$\\
\hline
\end{tabular}\label{table:prior_elicitation}
\caption{Summary of prior elicitations used in this paper for the spike-and-slab prior and Beta-Binomial prior.}
\end{table}
Table \ref{table:prior_elicitation} summarizes the choices of prior parameters used in this paper.
It should be noted that while these provide a reasonable trade-off  between including small effects and excluding spurious ones, it may be desirable to focus on one over the other. For instance, if there is no interest in including small effects, one could use a larger value for $C_{\beta}$ such as $1$ to improve specificity. The sensitivity analysis in Section \ref{sec:sens} shows that the prior elicitation is not particularly sensitive to moderate perturbations in the spike-and-slab parameters, so prioritizing minimizing false positives should be achieved by setting the slab variance substantially larger.

\textbf{Other parameters:} We do not directly define priors on $\rho$ and $\sigma^{2}$, but indirectly through $\tilde{\rho}$ and $\tilde{\sigma}^{2}$. The prior on $\tilde{\sigma}^{2}$ is $\text{IG}(c,d)$, with $c, d > 0$ positive constants, and the prior on $\tilde{\rho}$ is $N(0,\tau \tilde{\sigma}^{2})$ where $\tau > 0$ is a positive constant. We let the prior on $\tilde{\rho}$ depend on $\tilde{\sigma}^{2}$ so that the induced prior on $\rho$ has its density approach zero as $\rho \rightarrow \pm 1$. 
Choosing $c = d = 1$ leads to an induced prior on $\sigma^{2}$ with similar behaviour to an Inverse Gamma. The choice of $\tau$ in the prior for $\tilde{\rho}$ has little impact on the posterior inference for $\bm{\gamma^{S}},\bm{\gamma^{O}},\alpha_{0},\beta_{0},\bm{\alpha},\bm{\beta}$ and $\sigma^{2}$. However, it does have an impact on the posterior inference of $\rho$. The induced prior on $\rho$ depends on the choice of $\tau$: for small values such as $\tau = 0.5$, it behaves similarly to a uniform distribution, while for larger values like $\tau = 5$, it behaves similarly to a $Beta(1/2,1/2)$ distribution. Empirically we found that $\tau=5$ leads to more accurate posterior medians of $\rho$ than $\tau = 0.5$. See \ref{sec:rhoprior} for a further discussion.

\section{Gibbs samplers}\label{sec:gibbs}
In this section, we formulate Gibbs samplers for the sample selection model \eqref{eq:twoeqsampsel}
coupled with the prior structures proposed in Section \ref{sec:priors}. After initializing all model parameters, 
\begin{itemize}
\item[\textbf{Step 1.}] For $i = 1, \ldots, n$, sample from $\bm{s^{*}}$, where $s^{*}_{i} \mid \{y_{i}, s_{i}, \alpha_{0}, \beta_{0}, \bm{\alpha}, \bm{\beta}, \tilde{\rho}, \tilde{\sigma}\}$ is distributed according to
\begin{eqnarray*}
\begin{cases}
\mathcal{TN}_{(-\infty, 0)}(\alpha_{0} + \bm{w}_{i}^{\top}\bm{\alpha}, 1), & \text{if } y_{i} \text{ is missing},\\
\mathcal{TN}_{(0,\infty)}\left(\alpha_{0} + \bm{w}_{i}^{\top}\bm{\alpha} + \frac{\tilde{\rho}}{\tilde{\rho}^{2} + \tilde{\sigma}^{2}}(y_{i} - \beta_{0} - \bm{x}_{i}^{\top}\bm{\beta}), \frac{\tilde{\sigma}^{2}}{\tilde{\rho}^{2} + \tilde{\sigma}^{2}}\right), & \text{otherwise},
\end{cases}
\end{eqnarray*}
where $\mathcal{TN}_{(a,b)}$ denotes a truncated normal distribution on $(a,b)$.

\item[\textbf{Step 2.}] Sample $(\alpha_{0},\bm{\alpha}^{\top})^{\top}$ from $N(\bm{a^{*}}, \bm{A^{*}})$. 
\item[\textbf{Step 3.}] Jointly sample $(\beta_{0}, \bm{\beta}^{\top}, \tilde{\rho})^{\top}$ from $N(\bm{b^{*}}, \bm{B^{*}})$. 
\item[\textbf{Step 4.}] Sample $\tilde{\sigma}^{2}$ from $\text{IG}(c^{*},d^{*})$, where
\begin{align*}
c^{*} &= c + \frac{1}{2}\left(1 + \sum_{i=1}^{n} s_{i}\right) \, ,\\
d^{*} &= d + \frac{\tilde{\rho}^{2}}{2\tau} + \frac{1}{2}\sum_{\{i: s_{i}=1\}}\left(y_{i} - \beta_{0} -\bm{x}_{i}^{\top}\bm{\beta} - \tilde{\rho}(s^{*}_{i} - \alpha_{0} - \bm{w}_{i}^{\top}\bm{\alpha})\right)^{2} \, .
\end{align*}
\item[\textbf{Step 5.}] For $j = 1, \ldots, p$, sample $\gamma^{O}_{j} \mid \{\beta_{j}, v^{O}_{j}\} \sim Ber(r^{O}_{j})$ where
\begin{equation*}
r^{O}_{j} = \frac{r\phi\left(\beta_{j}/\left(\tau_{1,\beta}\sqrt{v^{O}_{j}}\right)\right)\pi_{1,\beta}(v^{O}_{j})}{r\phi\left(\beta_{j}/\left(\tau_{1,\beta}\sqrt{v^{O}_{j}}\right)\right)\pi_{1,\beta}(v^{O}_{j}) + (1-r)\phi\left(\beta_{j}/\left(\tau_{0,\beta}\sqrt{v^{O}_{j}}\right)\right)\pi_{0,\beta}(v^{O}_{j})} \, .
\end{equation*}
\item[\textbf{Step 6.}] For $k = 1, \ldots, q$, sample $\gamma^{S}_{k} \mid \{\alpha_{k}, v^{S}_{k}\} \sim Ber(r^{S}_{k})$ where
\begin{equation*}
r^{S}_{k} = \frac{r\phi\left(\alpha_{k}/\left(\tau_{1,\alpha}\sqrt{v^{S}_{k}}\right)\right)\pi_{1,\alpha}(v^{S}_{k})}{r\phi\left(\alpha_{k}/\left(\tau_{1,\alpha}\sqrt{v^{S}_{k}}\right)\right)\pi_{1,\alpha}(v^{S}_{k}) + (1-r)\phi\left(\alpha_{k}/\left(\tau_{0,\alpha}\sqrt{v^{S}_{k}}\right)\right)\pi_{0,\alpha}(v^{S}_{k})} \, .
\end{equation*}
\item[\textbf{Step 7.}] Sample $r \mid \{\bm{\gamma^{O}}, \bm{\gamma^{S}}\} \sim Beta(a_{1},b_{1})$, where
\begin{align*}
a_{1} &= a_{0} + \sum_{j=1}^{p} \gamma^{O}_{j} + \sum_{k=1}^{q} \gamma^{S}_{k} \, ,\\
b_{1} &= b_{0} + p + q - \sum_{j=1}^{p} \gamma^{O}_{j} - \sum_{k=1}^{q} \gamma^{S}_{k} \, .
\end{align*}
\item[\textbf{Step 8.}] For $j = 1, \ldots, p$, sample $v^{O}_{j} \mid \{\gamma^{O}_{j}, \beta_{j}\}$ from
\begin{equation*}
p(v^{O}_{j} \mid \gamma^{O}_{j}, \beta_{j}) \propto (1-\gamma^{O}_{j})\phi\left(\frac{\beta_{j}}{\tau_{0,\beta}\sqrt{v^{O}_{j}}}\right)\pi_{0,\beta} (v^{O}_{j}) + \gamma^{O}_{j}\phi\left(\frac{\beta_{j}}{\tau_{1,\beta}\sqrt{v^{O}_{j}}}\right)\pi_{1,\beta} (v^{O}_{j}) \, ,
\end{equation*}
where $\pi_{0,\beta}$ and $\pi_{1,\beta}$ are the mixing distributions.
\item[\textbf{Step 9.}] For $k = 1, \ldots, q$, sample $v^{S}_{k} \mid \{\gamma^{S}_{k}, \alpha_{k}\}$ from
\begin{equation*}
p(v^{S}_{k} \mid \gamma^{S}_{k}, \alpha_{k}) \propto (1-\gamma^{S}_{k})\phi\left(\frac{\alpha_{k}}{\tau_{0,\alpha}\sqrt{v^{S}_{k}}}\right)\pi_{0,\alpha} (v^{S}_{k}) + \gamma^{S}_{k}\phi\left(\frac{\alpha_{k}}{\tau_{1,\alpha}\sqrt{v^{S}_{k}}}\right)\pi_{1,\alpha} (v^{S}_{k}) \, ;
\end{equation*}
where $\pi_{0,\alpha}$ and $\pi_{1,\alpha}$ are the mixing distributions.

\item[\textbf{Step 10.}] Sample $v^{O}_{0} \sim \phi(\beta_{0}/(\sqrt{\eta^{O}v^{O}_{0}}))\pi_{I,\beta}(v^{O}_{0})$ and $v^{S}_{0} \sim \phi(\alpha_{0}/(\sqrt{\eta^{S}v^{S}_{0}}))\pi_{I,\alpha}(v^{S}_{0})$.
\end{itemize}
The exact expressions for $\bm{a^{*}}, \bm{A^{*}}, \bm{b^{*}}$ and $\bm{B^{*}}$ can be found in Appendix \ref{sec:vansampler}. Steps 1 through 7 have distributions that are well-known and can easily be sampled from. Steps 8, 9 and 10 depend on the choice of scale distributions $\pi_{0,\alpha}, \pi_{1,\alpha}, \pi_{0,\beta}$ and $\pi_{1,\beta}$. There is no closed form expression for a general scale mixing distribution, but certain choices lead to closed form sampling for these steps. A Laplace prior can be used by choosing the mixing distribution to be $\text{Exp}(1/2)$: in this case, the posterior distribution to sample from takes the form of an Inverse Gaussian (after conditioning in Step 8/9). A Student-$t$ prior can be used by choosing the mixing distribution to be Inverse Gamma, which is conjugate to the normal likelihood. In these cases and the normal case (by choosing the mixing distributions to be point masses at $1$), the sampler is in closed form.

The output of the above procedure will be a sample from the posterior $(\bm{\alpha}^{\top}, \bm{\beta}^{\top}, {\bm{\gamma^{S}}}^{\top}, {\bm{\gamma^{O}}}^{\top}, \tilde{\sigma}^{2}, \tilde{\rho})^{\top} \mid \{\bm{y}, \bm{s}\}$. While the sampler still relies on $(p+1) \times (p+1)$ and $(q+1) \times (q+1)$ matrix inversions, the dimensions are usually not so big in applications of interest that this is problematic.

\section{Simulation study}\label{sec:simulation}
In this section, we present a simulation study that aims at illustrating  how the prior calibration performs as the dimension increases in highly sparse situations, for small and moderately large samples and typical levels of missingness in the data. We compare the performance of the Gibbs samplers to alternatives such as Adaptive LASSO and stepwise selection.

\subsection{Simulation scenarios}
The true model is of the form in \eqref{eq:twoeqsampsel} with $\bm{\alpha} = \left(0.5, 1, 1.5, 0, \ldots, 0\right)/\sqrt{2} \in \mathbb{R}^{q}$, $\bm{\beta} = \left(0.25, 0.5, 1, 0, \ldots, 0\right) \in \mathbb{R}^{p}$, that is $3$ active variables and $p - 3$ spurious variables, and $p = q$. The effect sizes are chosen to represent a small, medium large effect respectively. The effect sizes for the probit equation are in line with \cite{certo::2016}. The intercept $\alpha_{0}$ is chosen dependent on the simulated covariates, so that the expected proportion of missing data is $0.3$. We choose $\beta_{0} = 0.5$ and $\sigma = 1$.
We generated $n$ covariates $\bm{w}_{i} \in \mathbb{R}^{q}$ such that their marginal distributions are standard normal and $\text{cov}(w_{j},w_{k}) = 0.5^{\lvert j - k \rvert}$. We let $\bm{w}_{i} = \bm{x}_{i}$ so that there is no exclusion restriction. For each $\bm{w}_{i}$, we generated values of $y_{i}$ and $s_{i}$, and repeated this for $1000$ Monte Carlo replicates (keeping the covariates fixed for each scenario).
We consider sample sizes $n = 500$ and $1000$, and dimensions $p = 10, 25$ and $50$,
We performed simulations for $\rho = 0, 0.3, 0.5$ and $0.7$, and found that the performance of each method was not dependent on $\rho$. For this reason, we only present the results for $\rho = 0.5$ in the main text, and include results for other values of $\rho$ in Appendix \ref{sec:corrs}.
We compare the following variable selection methods:
\begin{enumerate}
\item A normal-normal Class I spike-and-slab prior, with hyperparameters elicited as in Table \ref{table:prior_elicitation}, using Beta-Binomial hyperparameters $(1,1)$.
\item A Laplace-Laplace Class I spike-and-slab prior, with hyperparameters as in the normal-normal case, choosing the scale parameter of the Laplace distributions such that they have equal variances to the normal case.
\item Adaptive LASSO as in \cite{ogundimu2022regularization}.
\item A vanilla implementation of forward selection which starts from the null model and, at each step, fits the maximum likelihood estimates of the parameters and adds the variable which minimizes the Bayesian Information Criterion (BIC) to either the selection or outcome equation (whichever minimizes the BIC). 
\end{enumerate}

\subsection{Performance measures}
The evaluation metrics we use are: Sensitivity (TPR) - the proportion of active variables correctly identified; Specificity (TNR) - the proportion of inactive variables correctly identified; True model rate (TMR) - the proportion of replicates where the selected model is the true model; Model size - the average size of the selected model.
For spike-and-slab samplers, the ``selected model'' is the median model, that is the model such that a variable is included if and only if the proportion of sampled models a variable is included in, also known as the posterior inclusion probability (PIP) is greater than $0.5$. Formally, a variable $\beta_{j}$ is ``included'' if more than half of the posterior samples of $\gamma^{O}_{j}$ are equal to $1$, and similarly for $\alpha_{k}$. \cite{barbieri2004optimal} shows in the normal linear case that this is the optimal predictive model. 

We run the Gibbs samplers for $10,000$ iterations, discarding the first $1,250$ iterations as burn-in. These choices provide a compromise between convergence of the chain and computational efficiency for $1,000$ replicates in each simulation scenario. This is enough to stabilize the median model used to evaluate performance, but due to slow mixing of the chain (as further discussed in Appendix \ref{sec:mcmc}), in real applications we recommend running the chain for longer, \textit{e.g.}~$50,000$ iterations as in Section \ref{sec:applications}. The \texttt{sampleSelection} R package is used to choose initial values for the spike-and-slab samplers. That is, the parameter estimates from the maximum likelihood fit are used as the starting parameters, and the initial value of $\bm{\gamma}^{S}$ and $\bm{\gamma}^{O}$ are chosen such that a variable is included in the model if and only if its parameter estimate (under the maximum likelihood fit) is significant at a $5\%$ level. For both the normal and Laplace cases, we choose $\tau=5$ for the prior on $\tilde{\rho}$, $Beta(1,1)$ for the prior on $r$, and use the same distribution for the intercept as the rest of the model, with the chosen slab variance as its variance. For the normal case, we let $\tau_{1,\alpha} = \tau_{1,\beta} = 0.5$, and we divide these by $\sqrt{2}$ for the Laplace case.

\subsection{Convergence issues}\label{sec:convergence}
In some instances, the maximum likelihood estimates can fail to converge for sample selection models. This can occur due to the non-convexity in the likelihood surface, potentially leading to convergence to local maxima \citep{olsen:1982}, or issues with exclusion restrictions. While the model may be theoretically identifiable, in practice there is often practical non-identifiability for small sample sizes, large $\rho$ and large numbers of variables, especially in the presence of collinearity \citep{puhani:2000}. 

This is a particularly significant problem for the implementation of Adaptive LASSO as in \cite{ogundimu2022regularization}, which relies on a least-squares approximation of the likelihood of the \textit{full} model around the maximum likelihood estimator. If maximum likelihood estimation fails then the algorithm breaks down. In each simulation scenario, we produced simulation replicates for Adaptive LASSO until we had $1,000$ where the MLE of the full model converged. For $n=1,000$ few extra replicates were required, but for $n=500$ there were issues. For $\rho = 0$, 37.2\% of cases failed to converge; for $\rho = 0.3$, 51.8\% of cases failed; for $\rho = 0.5$, 71.3\% cases failed; and for $\rho = 0.7$, 91.8\% of cases failed.

In contrast, the spike-and-slab sampler can still be run in these scenarios. As such, we used the original $1,000$ replicates, including replicates where the MLE failed to converge, for the spike-and-slab simulations, and this should be taken into account when comparing the results. Our initialization in cases where the MLE failed to converge was $0$ for all model parameters aside from $\sigma$ which we set to $1$.

The stepwise estimator also exhibited occasional issues with convergence, but less often than Adaptive LASSO, due to starting from a null model and on average including less variables than the Adaptive LASSO.


\subsection{Results}

The results of the simulations can be found in Appendix \ref{sec:sim_main}. Table \ref{table:sim1} shows the performance of each method for $n=500$ and $p = 10, 25$ and $50$ respectively. It should be noted that the medium and large effects are included with probability close to one, so that the sensitivity is almost entirely determined by the inclusion of the small effect.
For $p=10$, the performance of the normal spike-and-slab sampler and the stepwise selection method are close. The normal spike-and-slab elicitation attains higher inclusion rates of small effects at the cost of lower specificity, while stepwise selection does the opposite. The Laplace spike-and-slab elicitation includes slightly more variables than the normal elicitation, while having lower specificity. The Adaptive LASSO performs similarly to stepwise in the outcome equation but struggles with false positives in the selection part of the equation, with an average model size of $3.458$.
We emphasize that these results for spike-and-slab priors depend on prior elicitation. For example, a larger slab variance for either prior distribution would improve specificity but exclude the small effect more often.

As the sparsity of the true model increases, the performance of every method deteriorates. The methods based on spike-and-slab priors exclude the small effect more often with increasing sparsity; this is a result of the multiple testing penalty imposed by the Beta-Binomial prior.
Both Adaptive LASSO and stepwise selection struggle particularly as the sparsity increases. For $p=50$, the model size of Adaptive LASSO in the selection equation averages $4.57$ when the true model size is $3$, and it increasingly excludes the small effect in the outcome equation as $p$ increases. Stepwise selection performs better but still has issues with false positives as $p$ increases, averaging model sizes in each equation greater than $3.4$ when $p=50$. Both spike-and-slab elicitations attain substantially larger true model rates and sensitivity than each of the Adaptive LASSO and stepwise methods for $p=25$ and $p=50$.

Table \ref{table:sim2} compares the methods for $n=1000$. All methods perform significantly better, with Adaptive LASSO producing a competitive performance at this sample size. Adaptive LASSO exhibits better performance for the outcome equation in low dimensions, while the spike-and-slab priors outperform Adaptive LASSO in the presence of sparsity when taking the selection equation into account. Stepwise selection still attains strong performance for $p=10$ but as with $n=500$, the performance is significantly worse as $p$ increases.


\section{Real data applications}\label{sec:applications}
In this section, we present two real data applications that illustrate the use of the proposed Bayesian variable selection methodology and how it compares against adaptive LASSO and stepwise selection. The first example analyzes ambulatory expenditures \citep{camer10}, whereas the second example analyzes data from the RAND Health Insurance Experiment (RAND HIE), a comprehensive study on the effect of health insurance on medical expenditures.
For these applications, we standardize all the covariates and examine the parameter estimates and standard deviations using the standardized data, as the true models are unknown. For the spike-and-slab methods, we additionally look at the posterior inclusion probabilities and use the same priors as in Table \ref{table:prior_elicitation}, using Beta-Binomial elicitation $(1,p+q)$ (to account for low sparsity). The samplers are run for $50,000$ iterations, with the first $5,000$ being burn-in. We report the posterior median of the parameters, based on the proposed spike-and-slab priors, as a Bayes' estimate.
We additionally report the results from fitting the full model using the \texttt{sampleSelection} R package. We present the results for Class II priors and additional results in the Appendix.
We briefly consider post-selection inference. To compare models without re-fitting the data, we follow the approach of \cite{liang2008mixtures} by comparing parameter posteriors conditional on the given models, \textit{i.e.}~$\left(\bm{\alpha}, \bm{\beta}, \sigma, \rho\right) \mid \bm{\gamma}$. In our case we consider the sub-sample where the sampled model is $\bm{\gamma}$ (and only considering parameters with $\gamma_{j} = 1$). We evaluate the performance of different models using the Bayesian leave-one-out estimate of the expected log pointwise predictive density, referred to as $elpd_{loo}$ \citep{vehtari2017practical}. We use the \texttt{loo} package in R to compute this.

\subsection{Ambulatory expenditures data}
The ambulatory expenditures data contains information about several explanatory variables, such as age, gender, education status (\texttt{educ}), ethnicity (\texttt{blhisp}), number of chronic conditions (\texttt{totchr}), insurance status (\texttt{ins}) and income. 
The main aim is to study the effect of these variables on ambulatory expenditures.
Not all patients had money spent, so that there are missing outcomes, and since we expect the decision to spend to be linked with the cost, this is a case of sample selection bias.
The dataset contains $n=3,328$ observations, of which $516$ ($15.8\%$) have missing expenditure. This dataset has been studied in previous publications \citep{marchenko2012heckman,ogundimu2022regularization}, and in line with these references we use log-expenditure (\texttt{lambexp}) as the outcome variable. 
We choose the same predictors as in previous publications, for comparison. That is, we let $\bm{x} = (\texttt{age}, \texttt{female}, \texttt{educ}, \texttt{blhisp}, \texttt{totchr}, \texttt{ins})^{\top}$ and $\bm{w} = (\bm{x}^{\top}, \texttt{income})^{\top}$ so that \texttt{income} is an exclusion restriction. The use of the \texttt{income} for this purpose is questionable \citep{camer10}, and an advantage of using variable selection methods in this context is to determine whether such restrictions are necessary.

Table \ref{table:amb} shows the results of applying each method to the data. The variables selected by the spike-and-slab model and stepwise are identical, and Adaptive LASSO selects similarly. All three methods exclude \texttt{educ} and \texttt{ins} from the outcome equation, with posterior inclusion probabilities being small for both variables. 
The major difference is in the exclusion restriction. The Adaptive LASSO still includes the exclusion restriction \texttt{income}, albeit shrunk more than the other included variables. 
The spike-and-slab normal, on the other hand, excludes \texttt{income} from the selection equation, with posterior inclusion probability $0.349$. It is still included in a considerable number of posterior samples, so that the posterior distribution is bimodal, but the median estimate of the parameter is very close to zero as it lies in the spike part of the posterior instead of the slab. 
The variable \texttt{ins} is also somewhat shrunk in the selection equation, with inclusion probability $0.571$. The larger standard deviations in each case is due to the bimodality of the posterior when the inclusion probability is close to 0.5.

\begin{table}[ht]
\centering
\begin{tabular}{|c|ccc|cc|cc|cc|}
\hline & \multicolumn{3}{|c|}{Spike-and-slab normal} & \multicolumn{2}{|c|}{ALASSO} & \multicolumn{2}{|c|}{Stepwise} & \multicolumn{2}{|c|}{Full model}\\
  & PIP & Est. & S.D. & Est. & S.D. & Est. & S.D. & Est. & S.D.\\
 \hline
 & \multicolumn{9}{c|}{Selection equation}\\
 \hline
(Intercept) & - & 1.276 & 0.038 & 1.270 & 0.038 & 1.278 & 0.038 & 1.283 & 0.038 \\

educ & 1.000 & 0.177 & 0.032 & 0.158 & 0.030 & 0.186 & 0.029 & 0.159 & 0.031 \\

age & 0.949 & 0.112 & 0.039 & 0.089 & 0.030 & 0.111 & 0.030 & 0.099 & 0.031 \\

income & 0.349 & 0.007 & 0.042 & 0.054 & 0.034 & 0.000 & 0.000 & 0.072 & 0.035 \\

female & 1.000 & 0.320 & 0.031 & 0.320 & 0.030 & 0.321 & 0.030 & 0.331 & 0.030 \\

totchr & 1.000 & 0.602 & 0.055 & 0.609 & 0.055 & 0.612 & 0.055 & 0.615 & 0.055 \\

blhisp &  1.000 & -0.171 &  0.029 & -0.161 &  0.028 & -0.170 &  0.028 & -0.168 &  0.029 \\

ins & 0.571 & 0.048 & 0.046 & 0.062 & 0.030 & 0.085 & 0.030 & 0.082 & 0.030 \\
 \hline
 & \multicolumn{9}{c|}{Outcome equation}\\
 \hline
(Intercept) & - & 6.563 & 0.060 & 6.585 & 0.054 & 6.545 & 0.055 & 6.514 & 0.055 \\

educ & 0.116 & 0.004 & 0.018 & 0.000 & 0.000 & 0.000 & 0.000 & 0.048 & 0.027 \\

age & 1.000 & 0.230 & 0.026 & 0.222 & 0.026 & 0.232 & 0.026 & 0.238 & 0.026 \\

female & 1.000 & 0.158 & 0.032 & 0.141 & 0.030 & 0.165 & 0.030 & 0.174 & 0.030 \\

totchr & 1.000 & 0.399 & 0.031 & 0.390 & 0.030 & 0.406 & 0.030 & 0.417 & 0.030 \\

blhisp &  0.895 & -0.094 &  0.039 & -0.076 &  0.028 & -0.103 &  0.028 & -0.101 &  0.028 \\

ins &  0.033 & -0.001 &  0.008 &  0.000 &  0.000 &  0.000 &  0.000 & -0.014 &  0.025 \\
 \hline
$\sigma$ & - & 1.286 & 0.024 & 1.281 & 0.024 & 1.277 & 0.021 & 1.271 & 0.018\\
 $\rho$ & - & -0.265 &  0.154 & -0.318 &  0.139 & -0.215 &  0.145 & -0.131 &  0.147 \\
\hline
\end{tabular}
\caption{Results from ambulatory data.
ALASSO refers to Adaptive LASSO, Stepwise refers to forward selection and ``Full model'' to
the full model fit using the \texttt{sampleSelection} package. ``PIP'' refers to posterior inclusion probabilities for each variable, ``Est.'' refers to parameter estimates (posterior medians for spike-and-slab) and ``S.D.'' refers to standard deviations of the parameter estimates.}\label{table:amb}
\end{table}

Another difference is in the estimation of $\rho$. The full model estimates a significantly smaller value of $\rho$ compared to the Adaptive LASSO estimates and posterior median of the spike-and-slab normal.
The spike-and-slab normal priors and Adaptive LASSO produce estimates of $\rho$ further away from zero.

For post-selection inference, we compared the spike-and-slab median model and Adaptive LASSO model using samples from the spike-and-slab sampler, and also include the model returning the lowest $elpd_{loo}$ of the $10$ most sampled models, with results reported in table \ref{table:ambpost}. The best model and Adaptive LASSO include more variables than spike-and-slab, but the gain in predictive performance is relatively small, suggesting these additional variables contribute only modestly to predictive power.

\begin{table}[ht]
\centering
\begin{tabular}{|c|cc|}
\hline
 & $elpd_{loo}$ & Model size \\
 \hline
Best model & -5086.0 & 12 \\
\hline
Adaptive LASSO model & -5087.5 & 11 \\
\hline
Spike-and-slab median model & -5087.7 & 10\\
\hline
\end{tabular}
\caption{Values of $elpd_{loo}$ of various models for the ambulatory data. The $elpd_{loo}$ for each model was computed using only parameter samples corresponding to iterations where the given model was sampled. The ``best model'' refers to the model with the lowest $elpd_{loo}$ of the  10 most sampled models by the spike-and-slab sampler.}\label{table:ambpost}
\end{table}

\subsection{RAND data}
We now analyze a data set from the RAND Health Insurance Experiment (RAND HIE). This is a comprehensive study from $1971$ to $1986$ of the impact of randomized health insurance on health care cost, utilization and outcomes. This data set was used by \cite{cameron2005microeconometrics} to analyze how the patient’s use of health services is affected by types of randomly assigned health insurance.
In line with previous applications, we use the variable \texttt{lnmeddol}, the logarithm of medical expenses per individual, as the outcome variable. We let $\bm{x}$ consist of the logarithm of coinsurance rate plus 1 ($\texttt{logc} = \log(\texttt{coins}+1)$), the dummy variable for individual deductible plan (\texttt{idp}), the logarithm of participation incentive payment (\texttt{lpi}), an artificial variable \texttt{fmde} that is $0$ if $\texttt{idp} = 1$ and $\log(\max(1,\texttt{mde}/(0.01*\texttt{coins}))$ otherwise (where \texttt{mde} is the maximum expenditure offer), physical limitations (\texttt{physlm}), the number of chronic diseases (\texttt{disea}), dummy variables for good (\texttt{hlthg}), fair (\texttt{hlthf}) and poor (\texttt{hlthp}) self-rated health (where the baseline is excellent self-rated health), the log of family income (\texttt{linc}), the log of family size (\texttt{lfam}), education of household head in years (\texttt{educdec}), age of individual in years (\texttt{xage}), a dummy variable for female individuals (\texttt{female}), a dummy variable for individuals younger than $18$ years (\texttt{child}), a dummy variable for female individuals younger than $18$ years (\texttt{fchild}), and a dummy variable for black household heads (\texttt{black}). The selection variable is \texttt{binexp} which indicates whether the medical expenses are positive, and let $\bm{w} = \bm{x}$ so that there is no exclusion restriction.
A subsample is selected so that the study year is $2$ and \texttt{educdec} is not ``NA''. There are $5,574$ observations, of which $1,293$ have zero medical expenses so that \texttt{lnmeddol} is missing.
The table \ref{table:rand} contains the results of each method applied to the subsample. All the methods estimate similar values for $\rho$ of around $0.73$, aside from stepwise selection which estimates $\rho = -0.225$. Stepwise selection also selects a completely different model, for instance including \texttt{idp} in the selection equation which is excluded from both spike-and-slab and Adaptive LASSO models, but excluding \texttt{logc} from the outcome equation when it has inclusion probability $1.000$ in the spike-and-slab model and a large coefficient in the Adaptive LASSO model. The model selected by stepwise selection attains higher BIC than the model selected by the spike-and-slab sampler when the maximum likelihood estimates are fit to both models, suggesting that the search method has not been able to identify the optimal model for this dataset.
The inference is otherwise similar between the normal spike-and-slab prior and Adaptive LASSO. In the selection equation, the same variables are excluded. The only difference in this regard is that the normal spike-and-slab prior excludes $\texttt{hlthg}$ from the outcome, with the inclusion probability being $0.217$, whereas the Adaptive LASSO does not exclude this variable, though it does heavily shrink it to almost the same extent as the normal spike-and-slab prior. We also observe that the normal spike-and-slab prior produces larger estimates of the coefficients than Adaptive LASSO does. 

\begin{figure}[h]
\includegraphics[width=\textwidth]{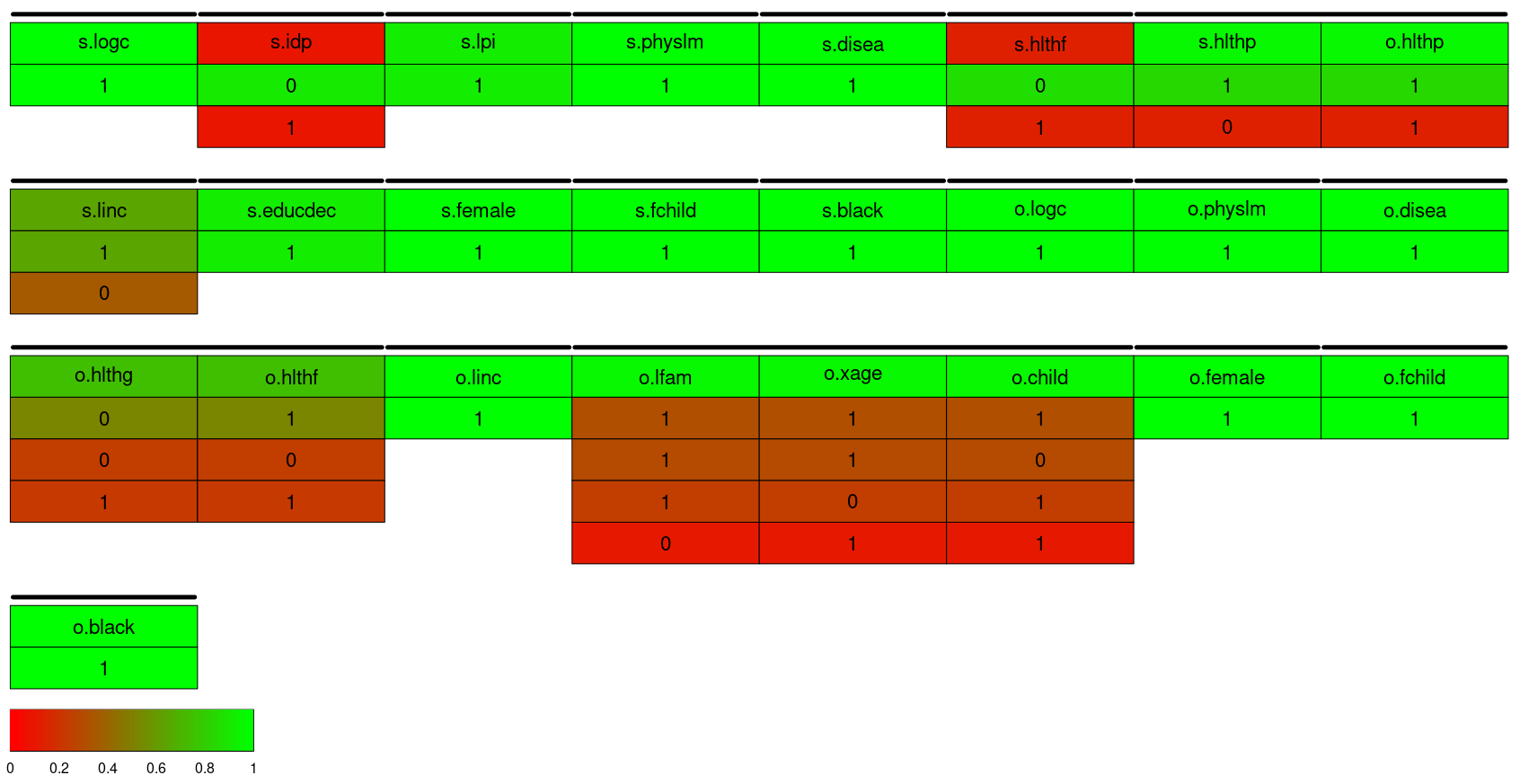}
\caption{Visualisation of a 50\% credible set for the model space posterior. The black bars above variables indicate groups - if a line is not broken between two variables, they are considered to be in the same group. The numbers indicate whether a variable is included or not and the color indicates the posterior inclusion probability of each listed group permutation.}\label{fig:rand_credible}
\end{figure}

We take advantage of recent advances in \cite{griffin2024expressing} to visualize the posterior in the model space. Figure \ref{fig:rand_credible} shows a 50\% credible set for the model space posterior. Of most interest from this visualisation is the relation of the group consisting of \texttt{lfam}, \texttt{xage} and \texttt{child}, in particular the latter two. \texttt{xage} and \texttt{child} have a correlation of $-0.804$, and from Figure \ref{fig:rand_credible}, it is clear that there is significant overlap in the information each variable brings into the model, as all three combinations involving one (or both) have similar posterior inclusion probability. In this case, both variables have inclusion probabilities greater than $0.5$, but if a higher cutoff were to be used, the visualisation suggests that at least one of these variables should be included regardless. It is also worth noting that \texttt{hlthg} and \texttt{hlthf} are grouped by this visualisation, which is where the only difference between the Adaptive LASSO model and the median model of the spike-and-slab occurs.

For post-selection inference, we compare the spike-and-slab median model and Adaptive LASSO model, using samples from the spike-and-slab sampler to obtain $elpd_{loo}$ values as in table \ref{table:randpost}. The best model  of the 20 most sampled models in this case coincides with the model chosen by Adaptive LASSO, which includes one additional variable, but the gain in predictive performance by including this variable is modest.

\begin{table}[ht]
\centering
\begin{tabular}{|c|cc|}
\hline
 & $elpd_{loo}$ & Model size \\
 \hline
Adaptive LASSO model (best model) & -8541.6 & 23 \\
\hline
Spike-and-slab median model & -8548.2 & 22\\
\hline 
\end{tabular}
\caption{Values of $elpd_{loo}$  of various models for the RAND data. The $elpd_{loo}$ for each model was computed using only parameter samples corresponding to iterations where the given model was sampled. The ``best model'' refers to the model with the lowest $elpd_{loo}$ of the  20 most sampled models by the spike-and-slab sampler.}\label{table:randpost}
\end{table}

\subsection{Sensitivity analysis}\label{sec:sens}
In this section, we perform a sensitivity analysis for the spike and slab variances on the two datasets.\\
Let $\tau_{1,\alpha}^{*} = \sqrt{3}/\pi$ and $\tau_{1,\beta}^{*} = (4\log(500))^{-1/2}(\log(n))^{1/2}$. Alongside those values, which are the defaults for the previous analysis, we consider two additional specifications. We consider $\tau_{1,\alpha} = 2\tau_{1,\alpha}^{*}$ and $\tau_{1,\beta} = 2\tau_{1,\beta}^{*}$, alongside the choices $\tau_{1,\alpha} = 0.5\tau_{1,\alpha}^{*}$ and $\tau_{1,\beta} = 0.5\tau_{1,\beta}^{*}$. In each case, the variances have been multiplied or shrunk by a factor of $4$, so that the values have change significantly. Other than the choice of spike-and-slab variances, we use the same elicitation as in the previous two sections.
Table \ref{table:amb_slab} shows the comparison between the three different choices on the ambulatory data. Similar tables for the spike variance and analogous tables for the RAND data can also be found in the table in Appendix \ref{sec:sens_tables}. As expected, the larger slab variances have smaller inclusion probabilities for every variable. Despite the difference in prior variances, though, the difference between the models is not drastic. The only difference in the maximum a posteriori model is the exclusion of \texttt{ins} in the selection equation by the larger slab variances. The posterior median of the coefficients for variables with inclusion probability not close to $1$ are shrunk more by the larger variances, but otherwise are similar.
Similar behaviour occurs for the RAND data (table \ref{table:rand_slab}). Due to the high collinearity between the predictors, some variables actually have higher inclusion probability under the larger slab variance, when highly correlated variables are excluded in their place.
Tables \ref{table:amb_spike} and \ref{table:rand_spike} show that posterior inference based on the spike-and-slab prior is less sensitive to the spike variance than the slab variance. There is little difference in inclusion probability for most variables.
\section{Conclusion}\label{sec:discussion}
We proposed two continuous spike-and-slab prior structures for variable selection in the context of sample selection models, and provided practical guidelines for calibrating these priors.
We developed Gibbs samplers, with closed-form tractable conditionals, for sample selection models coupled with the proposed prior structures. 
An appealing feature of the proposed prior structures and Gibbs samplers is that they allow for using (different) scale mixture of normals for the spike and slab components. 
We have shown that these Gibbs samplers are scalable to the dimensions of interest in practice. The time complexity is not linear, as clearly seen in Section \ref{sec:comptimes}, so it will not scale well to dimensions $p > > 200$. However, applications of interest studied in sample selection literature are seldom higher dimensional than this.

Our simulation study shows that the proposed Bayesian variable selection methodology offers a good performance in terms of sensitivity and specificity of the selected variables. Moreover, it exhibits similar performance to Adaptive LASSO in low dimensions and large sample sizes, while identifying the true model more frequently for extremely sparse models and smaller sample sizes.
A significant advantage of our proposed methodology is that, unlike Adaptive LASSO, our proposed methodology does not suffer from convergence issues for moderately small sample sizes and higher dimensions.


There exist several natural extensions of our work. 
It is desirable to extend theoretical results, similar to \cite{narisetty2014bayesian}, to sample selection models, and to use these results to inform prior elicitation, or use a data-driven approach to elicitate the hyperparameters instead \citep{tadesse2021handbook}.
Regarding scalability, one could extend the Skinny Gibbs algorithm \citep{narisetty:2019} to sample selection, which would let the algorithm scale to higher dimensions, or to develop \textit{ad hoc} variational Bayes methods \citep{tadesse2021handbook}.
A similar approach could be taken with EM algorithms, such as the spike-and-slab LASSO by \cite{rovckova2018spike}. 
We assumed bivariate normal errors in the two-equation model \eqref{eq:twoeqsampsel}. It would be possible to extend this model to non-normal errors, such as scale mixtures of normals (or other distributions with a tractable stochastic representation), which would allow to use the Gibbs samplers developed in Section \ref{sec:gibbs} with additional steps. The proposed Gibbs samplers can also be extended to the case with binary outcomes, by using the stochastic representation of probit models through latent variables.  
The focus in this paper is primarily on model selection and the problem of post-selection inference requires further investigation. Because the continuous spike-and-slab returns posterior probabilities for each model, an extension of the methodology in this paper could incorporate Bayesian model averaging \citep{raftery1997bayesian, steel2020model} into the output of the sampler directly. The issue of excluded parameters having positive posterior density away from zero under continuous spike-and-slab priors would need to be carefully considered in this context.

\section*{Acknowledgements}
Adam Iqbal was supported by the Heilbronn Institute for Mathematical Research via funding from the EPSRC grant ``Additional Funding Programme for Mathematical Sciences'' (EP/V521917/1).

\bibliographystyle{unsrtnat}
\bibliography{references} 

\clearpage

\appendix
\section{Computational times}\label{sec:comptimes}
To test computational times for the spike-and-slab Gibbs sampler, we recorded the running time of the method on one replicate from the simulation study. That is, we simulated $\bm{x}_{i} = (x_{i,1}, \ldots, x_{i,p})^{\top}$ as i.i.d. normal for $i = 1, \ldots, n$, let $\bm{w} = \bm{x}$, then generated bivariate normal errors to obtain $(y_{i}, s_{i})^{\top}$ from \eqref{eq:twoeqsampsel}, with $\sigma=1$. We chose $\rho = 0.5$ and to avoid convergence issues, $n=1000$. For the spike-and-slab normal, the chain was run for $10,000$ iterations, discarding the first $1,250$ as burn-in, as in the simulation study. We varied $p$ from $25$ to $200$ (and let $q = p$) to see how the running times scale to very high dimensions. The computations were performed on a computer with 16GB of RAM and an AMD Ryzen 5 3600 GPU. The reason this was run separately from the rest of the simulation study is because the original simulations were run in parallel and only considered up to $p=50$.
\begin{table}[ht]
\centering
\begin{tabular}{|c|c|c|c|c|}
\hline
Method & $p=25$ & $p=50$ & $p=100$ & $p=200$ \\
\hline
Spike-and-slab normal & 15.5 & 39.3 & 130.7 & 609.0 \\
\hline
\end{tabular}
\caption{Computational times for the spike-and-slab prior, for $n=1000$. All times are in seconds.} \label{table:times}
\end{table}

\section{Induced prior for $\rho$}\label{sec:rhoprior}

\begin{figure}[ht]
\centering
\includegraphics[width=0.75\textwidth]{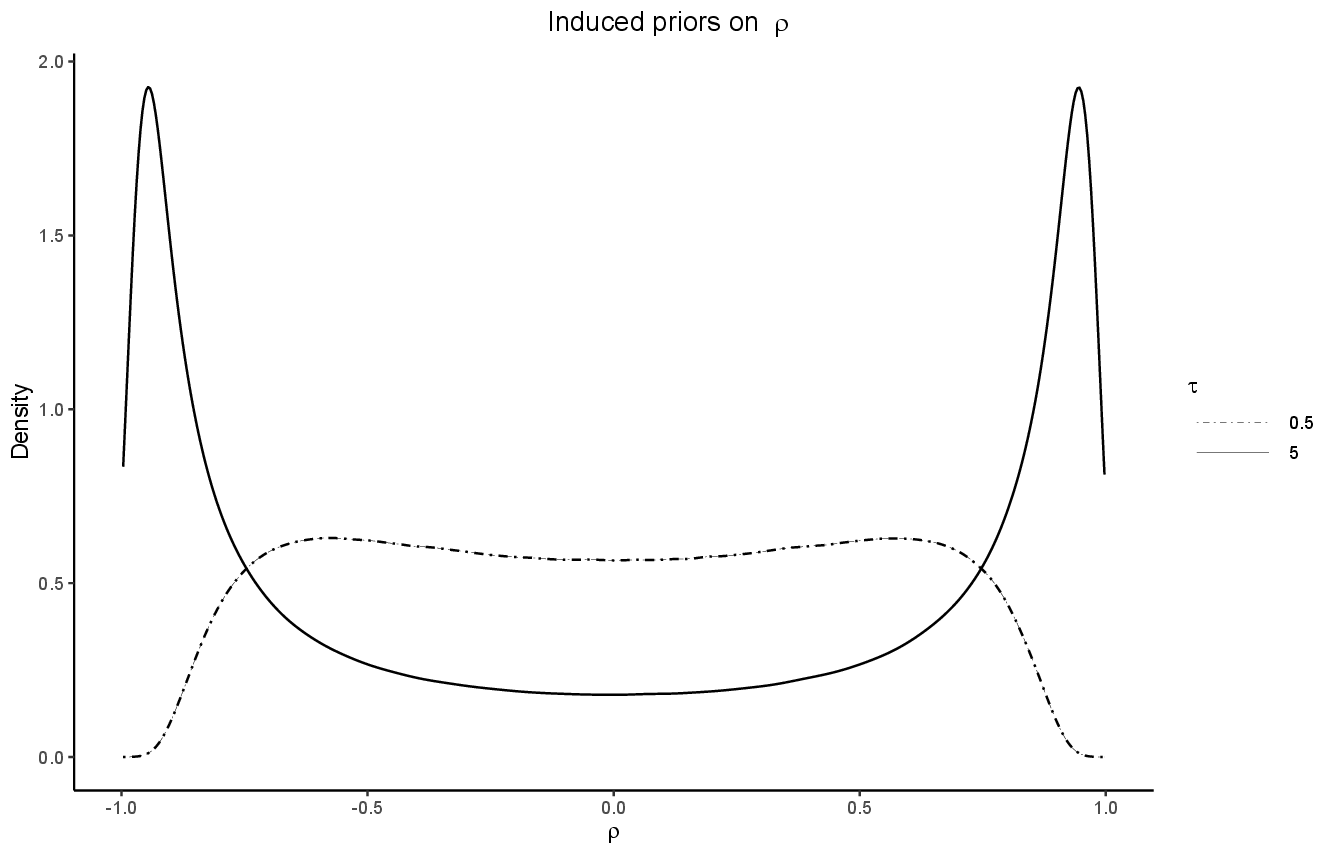}
\caption{The induced priors on $\rho$ for $\tau = 5$ and $0.5$ respectively. Smaller choices of $\tau$ are closer to a uniform distribution, while larger choices resemble Beta distributions, except approaching zero at the tails.}\label{fig:rhoprior}
\end{figure}

Recall that $\tilde{\sigma}^{2} \sim IG(c,d)$ and $(\tilde{\rho} \mid \tilde{\sigma}^{2}) \sim N(0, \tau\tilde{\sigma}^{2})$, with $c, d, \tau > 0$ positive constants. $\tau$ controls the behaviour of the induced prior on $\rho$. Figure \ref{fig:rhoprior} shows the shape of two induced priors on $(0,1)$, one with $\tau = 0.5$ and the other with $\tau = 5$. The case $\tau=5$ closely resembles a $Beta(1/2,1/2)$ prior, while $\tau=0.5$ resembles a uniform prior, aside from both approaching zero at the bounds $0$ and $1$. 

Alongside the simulation study in Section \ref{sec:simulation}, we ran the same simulations for the normal-normal spike-and-slab sampler with $\tau = 0.5$ (and otherwise the same choices of prior parameters). Table \ref{table:rho_comparison} shows how the posterior median of $\rho$ differed between these two priors. Both priors underestimate $\rho$, but the prior with $\tau = 5$ consistently gives estimates closer to the true value. The posterior inference for other parameters (not shown here) is similar for both choices of $\tau$, so we prefer $\tau = 5$ to $0.5$.

\begin{table}[ht]
\centering
\begin{tabular}{|cc|cc|}
\hline
 
  n & True $\rho$ & $\tau=0.5$ & $\tau=5$ \\
\hline
500 & 0 & -0.112 & -0.128\\
500 & 0.3 & 0.132 & 0.148\\
500 & 0.5 & 0.323 & 0.363\\
500 & 0.7 & 0.558 & 0.615\\
\hline
1000 & 0 & -0.028 & -0.030\\
1000 & 0.3 & 0.246 & 0.259\\
1000 & 0.5 & 0.442 & 0.464\\
1000 & 0.7 & 0.647 & 0.674\\
\hline
\end{tabular}
\caption{Comparison of posterior median of $\rho$ for $\tau=0.5$ and $\tau=5$. Uses spike-and-slab normal priors for each case, with the same elicitation and same $1000$ replicates as in the simulation study. The posterior medians of $\rho$ are very similar for all $p$, so we only include $p=10$ for brevity.}\label{table:rho_comparison}
\end{table}

\clearpage
\section{Gibbs sampler for sample selection models}\label{sec:vansampler}

Here, we reproduce the Gibbs sampler from \citep{van2011bayesian} for completeness. Let the priors on the parameters be
\begin{align*}
(\alpha_{0},\bm{\alpha}^{\top})^{\top} &\sim N(\bm{a},\bm{A}) \, ,\\
(\beta_{0},\bm{\beta}^{\top})^{\top} &\sim N(\bm{b}, \bm{B}) \, ,\\
\tilde{\rho} \mid \tilde{\sigma} &\sim N(0, \tau \tilde{\sigma}^{2}) \, ,\\
\tilde{\sigma}^{2} &\sim \text{IG}(c,d) \, .
\end{align*}
where $\bm{a} \in \mathbb{R}^{p+1}$, $\bm{A} \in \mathbb{R}^{(p+1)\times(p+1)}$, $\bm{b} \in \mathbb{R}^{q+1}$, $\bm{B} \in \mathbb{R}^{(q+1)\times(q+1)}$, $\tau \in (0,1)$ and $c, d > 0$ are real constants.\\
Further define $\bm{X} = (\bm{1},\bm{x}_{1},\ldots,\bm{x}_{n})$ and $\bm{W} = (\bm{1},\bm{w}_{1},\ldots,\bm{w}_{n})$.
Let $\bm{W}_{0}$ be the design matrix $\bm{W}$ with only the observations that have missing $y_{i}$, and $\bm{W}_{1}$ the design matrix with observations that have non-missing $y_{i}$. Similarly, let $\bm{X}_{1}$ be the design matrix $\bm{X}$ with the observations that have non-missing $y_{i}$, $\bm{y}_{1}$ be the non-missing outcomes, $\bm{s^{*}}_{1}$ the corresponding values of $s^{*}$ and $\bm{s^{*}}_{0}$ the values of $s^{*}$ for observations with missing outcomes. Furthermore, let $\bm{g} = (\bm{\beta}^{\top}, 0)^{\top}$ and 
\begin{eqnarray*}
\bm{G} =
\begin{pmatrix}
\bm{B} & \bm{0}\\
\bm{0} & \tau \tilde{\sigma}^{2}
\end{pmatrix} \, ,
\end{eqnarray*}
and let $\bm{Z}$ be the design matrix $\bm{X}_{1}$ with an additional column $\bm{s^{*}}_{1} - \bm{W}_{1}\bm{\alpha}$. Then the Gibbs sampler as in \cite{van2011bayesian} is as follows:
\begin{enumerate}
\item Sample from $\bm{s^{*}}$, where for $i = 1, \ldots, n$ where $s^{*}_{i} \mid \{y_{i}, s_{i}, \alpha_{0}, \beta_{0}, \bm{\alpha}, \bm{\beta}, \tilde{\rho}, \tilde{\sigma}\}$ is distributed according to
\begin{eqnarray*}
\begin{cases}
\mathcal{TN}_{(-\infty, 0)}(\alpha_{0} + \bm{w}_{i}^{\top}\bm{\alpha}, 1) & \text{if } y_{i} \text{ is missing},\\
\mathcal{TN}_{(0,\infty)}\left(\alpha_{0} + \bm{w}_{i}^{\top}\bm{\alpha} + \frac{\tilde{\rho}}{\tilde{\rho}^{2} + \tilde{\sigma}^{2}}(y_{i} - \beta_{0} - \bm{x}_{i}^{\top}\bm{\beta}), \frac{\tilde{\sigma}^{2}}{\tilde{\rho}^{2} + \tilde{\sigma}^{2}}\right)& \text{otherwise}
\end{cases}
\end{eqnarray*}
where $\mathcal{TN}_{(a,b)}$ denotes a truncated normal distribution on $(a,b)$.
\item Sample $(\alpha_{0},\bm{\alpha}^{\top})^{\top}$ from $N(\bm{a^{*}}, \bm{A^{*}})$, where
\begin{align*}
\bm{A^{*}} &= \left(\bm{A}^{-1} + \bm{W}_{0}^{\top}\bm{W}_{0} + \left(\frac{\tilde{\sigma}^{2}}{\tilde{\sigma}^{2} + \tilde{\rho}^{2}}\right)^{-1}\bm{W}_{1}^{\top}\bm{W}_{1}\right)^{-1} \, ,\\
\bm{a^{*}} &= \bm{A^{*}}\left(\bm{A}^{-1}\bm{a} + \bm{W}_{0}^{\top}\bm{s^{*}}_{0} + \left(\frac{\tilde{\sigma}^{2}}{\tilde{\sigma}^{2} + \tilde{\rho}^{2}}\right)^{-1}\bm{W}_{1}^{\top}\left(\bm{s^{*}}_{1} - \frac{\tilde{\rho}}{\tilde{\sigma}^{2} + \tilde{\rho}^{2}}(\bm{y}_{1} - \bm{X}_{1}\bm{\beta})\right)\right) \, .
\end{align*}
\item Jointly sample $(\beta_{0},\bm{\beta}^{\top}, \tilde{\rho})^{\top}$ from $N(\bm{b^{*}}, \bm{B^{*}})$, where
\begin{align*}
\bm{B^{*}} &= \left(\bm{G}^{-1} + \frac{1}{\tilde{\sigma}^{2}}\bm{Z}^{\top}\bm{Z}\right)^{-1} \, ,\\
\bm{b^{*}} &= \bm{B^{*}}\left(\bm{G}^{-1}\bm{g} + \frac{1}{\tilde{\sigma}^{2}}\bm{Z}^{\top}\bm{y}_{1}\right) \, .
\end{align*}
\item Sample $\tilde{\sigma}^{2}$ from $\text{IG}(c^{*},d^{*})$, where
\begin{align*}
c^{*} &= c + \frac{1}{2}\left(1 + \sum_{i=1}^{n} s_{i}\right) \, ,\\
d^{*} &= d + \frac{\tilde{\rho}^{2}}{2\tau} + \frac{1}{2}\sum_{\{i: s_{i}=1\}}\left(y_{i} - \beta_{0} - \bm{x}_{i}{^\top}\bm{\beta} - \tilde{\rho}(s^{*}_{i} - \alpha_{0} -\bm{w}_{i}^{\top}\bm{\alpha})\right)^{2} \, .
\end{align*}
\end{enumerate}


\clearpage
\section{Further details on the Gibbs samplers}

\subsection{Class I prior}
The prior variance matrix for the outcome equation, $\bm{B}$, corresponds to the $(p+1) \times (p+1)$ diagonal matrix  where $B_{1,1} = \eta^{O} v^{O}_{0}$ and $B_{j,j} = \left((1 - \gamma^{O}_{j-1})\tau_{0,\beta}^{2}v + \gamma^{O}_{j-1}\tau_{1,\beta}^{2}\right)v^{O}_{j-1}$ (with the mixing variable $v^{O}_{j-1}$ appearing to allow for scale mixture of normal priors). $\bm{A}$ is defined similarly.
Define $\bm{W}_{0}, \bm{W}_{1}, \bm{s^{*}}_{0}, \bm{s^{*}}_{1}, \bm{y}_{1}, \bm{g}, \bm{G}$ and $\bm{Z}$ as in Section \ref{sec:vansampler}.
Then we have that
\begin{align*}
\bm{A^{*}} &= \left(\bm{A}^{-1} + \bm{W}_{0}^{\top}\bm{W}_{0} + \left(\frac{\tilde{\sigma}^{2}}{\tilde{\sigma}^{2} + \tilde{\rho}^{2}}\right)^{-1}\bm{W}_{1}^{\top}\bm{W}_{1}\right)^{-1} \, ,\\
\bm{a^{*}} &= \bm{A^{*}}\left(\bm{W}_{0}^{\top}\bm{s^{*}}_{0} + \left(\frac{\tilde{\sigma}^{2}}{\tilde{\sigma}^{2} + \tilde{\rho}^{2}}\right)^{-1}\bm{W}_{1}^{\top}\left(\bm{s^{*}}_{1} - \frac{\tilde{\rho}}{\tilde{\sigma}^{2} + \tilde{\rho}^{2}}(\bm{y} - \bm{X}\bm{\beta})\right)\right) \, ,\\
\bm{B^{*}} &= \left(\bm{G}^{-1} + \frac{1}{\tilde{\sigma}^{2}}\bm{Z}^{\top}\bm{Z}\right)^{-1} \, ,\\
\bm{b^{*}} &= \bm{B^{*}}\left(\frac{1}{\tilde{\sigma}^{2}}\bm{Z}^{\top}\bm{y}_{1}\right) \, .
\end{align*}

\subsection{Class II prior}
The Gibbs sampler for the Class II prior is similar to the Class I prior, with only two differences. The first difference is that wherever $\tau_{0,\beta}$ and $\tau_{1,\beta}$ appear, $\tau_{0,\beta}\tilde{\sigma}$ and $\tau_{1,\beta}\tilde{\sigma}$ appear instead. The second difference are the values of $c^{*}$ and $d^{*}$ in sampling $\tilde{\sigma}^{2}$ (in Step 4 of the algorithm). That is, we instead have that
\begin{align*}
c^{*} &= c + \frac{1}{2}\left(1 + \sum_{i=1}^{n} s_{i}\right) + \frac{p+1}{2}\, ,\\
d^{*} &= d + \frac{\tilde{\rho}^{2}}{2\tau} + \frac{1}{2}\sum_{\{i: s_{i}=1\}}\left(y_{i} - \beta_{0} - \bm{x}_{i}{^\top}\bm{\beta} - \tilde{\rho}(s^{*}_{i} - \alpha_{0} - \bm{w}_{i}^{\top}\bm{\alpha})^{2}\right) + \frac{\beta_{0}^{2}}{2\eta^{O} v^{O}_{0}} \\
&+ \sum_{j=1}^{p} \frac{\beta_{j}^{2}}{2v^{O}_{j}((1-\gamma^{O}_{j})\tau_{0,\beta}^{2} + \gamma^{O}_{j}\tau_{1,\beta}^{2})}\, .
\end{align*}

\clearpage
\section{RAND data table}\label{sec:randtable}

\begin{table}[ht]
\centering
\resizebox{0.9 \textwidth}{!}{
\begin{tabular}{|c|ccc|cc|cc|cc|}
\hline & \multicolumn{3}{|c|}{Spike-and-slab normal} & \multicolumn{2}{|c|}{ALASSO} & \multicolumn{2}{|c|}{Stepwise} & \multicolumn{2}{|c|}{Full model}\\
 & PIP & Est. & S.D. & Est. & S.D. & Est. & S.D. & Est. & S.D.\\
 \hline
 & \multicolumn{9}{c|}{Selection equation}\\
 \hline
(Intercept) & - & 0.813 & 0.020 & 0.801 & 0.020 & 0.819 & 0.020 & 0.815 & 0.020 \\

logc &  1.000 & -0.229 &  0.030 & -0.198 &  0.022 & -0.218 &  0.023 & -0.218 &  0.054 \\

idp &  0.094 & -0.001 &  0.012 &  0.000 &  0.000 & -0.062 &  0.020 & -0.048 &  0.022 \\

lpi & 0.919 & 0.072 & 0.028 & 0.043 & 0.020 & 0.079 & 0.022 & 0.079 & 0.023 \\

fmde & 0.090 & 0.001 & 0.023 & 0.000 & 0.000 & 0.000 & 0.000 & 0.003 & 0.055 \\

physlm & 0.998 & 0.099 & 0.024 & 0.073 & 0.022 & 0.100 & 0.023 & 0.092 & 0.023 \\

disea & 1.000 & 0.154 & 0.023 & 0.150 & 0.023 & 0.169 & 0.024 & 0.143 & 0.024 \\

hlthg & 0.020 & 0.000 & 0.004 & 0.000 & 0.000 & 0.000 & 0.000 & 0.028 & 0.021 \\

hlthf & 0.131 & 0.001 & 0.018 & 0.000 & 0.000 & 0.000 & 0.000 & 0.060 & 0.022 \\

hlthp & 0.844 & 0.077 & 0.038 & 0.053 & 0.023 & 0.000 & 0.000 & 0.099 & 0.025 \\

linc & 0.640 & 0.048 & 0.034 & 0.024 & 0.020 & 0.052 & 0.020 & 0.068 & 0.020 \\

lfam &  0.019 &  0.000 &  0.005 &  0.000 &  0.000 &  0.000 &  0.000 & -0.017 &  0.022 \\

educdec & 0.922 & 0.071 & 0.027 & 0.050 & 0.019 & 0.087 & 0.021 & 0.089 & 0.021 \\

xage &  0.031 & 0.000 &  0.007 &  0.000 &  0.000 &  0.000 &  0.000 & -0.010 &  0.035 \\

female & 1.000 & 0.196 & 0.024 & 0.164 & 0.023 & 0.209 & 0.024 & 0.205 & 0.027 \\

child & 0.057 & 0.001 & 0.011 & 0.000 & 0.000 & 0.000 & 0.000 & 0.026 & 0.039 \\

fchild &  1.000 & -0.143 &  0.024 & -0.111 &  0.022 & -0.154 & 0.023 & -0.157 &  0.031 \\

black &  1.000 & -0.227 &  0.021 & -0.223 &  0.019 & -0.230 & 0.019 & -0.225 &  0.020 \\
 \hline
 & \multicolumn{9}{c|}{Outcome equation}\\
 \hline
(Intercept) & - & 3.550 & 0.039 & 3.561 & 0.038 & 4.124 & 0.096 & 3.543 & 0.036 \\

logc &  1.000 & -0.229 &  0.027 & -0.209 &  0.024 &  0.000 &  0.000 & -0.155 &  0.069 \\

idp &  0.033 & -0.001 &  0.007 &  0.000 &  0.000 & 0.000 & 0.000 & -0.066 &  0.029 \\

lpi & 0.046 & 0.001 & 0.010 & 0.000 & 0.000 & 0.000 & 0.000 & 0.040 & 0.028 \\

fmde &  0.037 & 0.000 &  0.011 &  0.000 &  0.000 & -0.116 & 0.022 & -0.081 &  0.067 \\

physlm & 1.000 & 0.124 & 0.025 & 0.102 & 0.024 & 0.080 & 0.023 & 0.115 & 0.024 \\

disea & 1.000 & 0.205 & 0.027 & 0.204 & 0.026 & 0.127 & 0.024 & 0.194 & 0.026 \\

hlthg & 0.236 & 0.002 & 0.026 & 0.026 & 0.023 & 0.000 & 0.000 & 0.075 & 0.025 \\

hlthf & 0.740 & 0.066 & 0.040 & 0.060 & 0.023 & 0.000 & 0.000 & 0.120 & 0.026 \\

hlthp & 0.964 & 0.100 & 0.031 & 0.086 & 0.023 & 0.085 & 0.021 & 0.124 & 0.023 \\

linc & 0.997 & 0.138 & 0.034 & 0.108 & 0.027 & 0.109 & 0.026 & 0.148 & 0.028 \\

lfam &  0.853 & -0.079 &  0.037 & -0.060 &  0.024 & -0.077 &  0.024 & -0.086 &  0.027 \\

educdec & 0.076 & 0.001 & 0.014 & 0.000 & 0.000 & 0.000 & 0.000 & 0.050 & 0.026 \\

xage & 0.749 & 0.115 & 0.071 & 0.100 & 0.037 & 0.104 & 0.037 & 0.096 & 0.041 \\

female & 1.000 & 0.279 & 0.035 & 0.252 & 0.031 & 0.148 & 0.030 & 0.275 & 0.032 \\

child &  0.670 & -0.110 &  0.083 & -0.109 &  0.043 & -0.136 &  0.044 & -0.097 &  0.048 \\

fchild &  1.000 & -0.234 &  0.044 & -0.204 &  0.037 & -0.122 &  0.036 & -0.224 &  0.039 \\

black &  1.000 & -0.205 &  0.030 & -0.197 &  0.029 & 0.000 &  0.000 & -0.207 &  0.029 \\
 \hline
$\sigma$ & - & 1.571 & 0.029 & 1.561 & 0.029 & 1.410 & 0.018 & 1.570 & 0.028\\
 $\rho$ & - & 0.729 &  0.039 &  0.721 &  0.038 & -0.225 & 0.076 & 0.736 & 0.034 \\
\hline
\end{tabular}}\label{table:rand}
\caption{Results from RAND health data.
ALASSO refers to Adaptive LASSO, Stepwise refers to forward selection and ``Full model'' to
the full model fit using the \texttt{sampleSelection} package.}
\end{table}
\clearpage

\clearpage

\section{Main simulation tables}\label{sec:sim_main}

\subsection{$n=500$}
\begin{table}[ht]
\centering
\begin{tabular}{|c|cccc|cccc|}
\hline
 
  \multirow{2}{4em}{Method} & \multicolumn{4}{c|}{Selection equation} & \multicolumn{4}{c|}{Outcome equation} \\

   & TMR & Size & Sens. & Spec. & TMR & Size & Sens. & Spec. \\
\hline
 & \multicolumn{8}{c|}{$p=10$} \\
\hline
Normal & 0.783 & 3.173 & 0.985 & 0.969 & 0.794 & 3.045 & 0.970 & 0.981\\
Laplace & 0.669 & 3.402 & 0.991 & 0.939 & 0.730 & 3.219 & 0.982 & 0.961\\
Stepwise$^{*}$ & 0.831 & 2.990 & 0.968 & 0.988 & 0.808 & 2.957 & 0.960 & 0.989\\
ALASSO$^{*}$ & 0.558 & 3.458 & 0.982 & 0.927 & 0.762 & 2.919 & 0.945 & 0.988\\
\hline
 & \multicolumn{8}{c|}{$p=25$} \\
\hline
Normal & 0.767 & 3.043 & 0.964 & 0.993 & 0.777 & 2.938 & 0.950 & 0.996\\
Laplace & 0.728 & 3.139 & 0.971 & 0.990 & 0.755 & 3.026 & 0.959 & 0.993\\
Stepwise$^{*}$ & 0.682 & 3.198 & 0.970 & 0.987 & 0.699 & 3.176 & 0.969 & 0.988\\
ALASSO$^{*}$ & 0.420 & 3.900 & 0.980 & 0.956 & 0.702 & 2.874 & 0.927 & 0.996\\
\hline
 & \multicolumn{8}{c|}{$p=50$} \\
\hline
Normal & 0.723 & 2.942 & 0.940 & 0.997 & 0.657 & 2.805 & 0.906 & 0.998\\
Laplace & 0.717 & 3.007 & 0.948 & 0.997 & 0.685 & 2.888 & 0.924 & 0.998\\
Stepwise$^{*}$ & 0.511 & 3.498 & 0.968 & 0.987 & 0.534 & 3.429 & 0.964 & 0.989\\ 
ALASSO$^{*}$ & 0.271 & 4.570 & 0.973 & 0.965 & 0.537 & 2.717 & 0.871 & 0.998\\
\hline
\end{tabular}
\caption{Evaluation metrics for $n=500, \rho=0.5$. ``TMR'' refers to the proportion of seeds where the true model was selected and ``Size'' to the number of variables in the model. Stepwise uses forward selection and ALASSO refers to Adaptive LASSO.\\
$^{*}$ Performance measures for Adaptive LASSO and stepwise selection were only evaluated over replicates that converged and had finite variance.}
\label{table:sim1}
\end{table}
\subsection{$n=1000$}

\begin{table}[ht]
\centering
\begin{tabular}{|c|cccc|cccc|}
\hline
 
  \multirow{2}{4em}{Method} & \multicolumn{4}{c|}{Selection equation} & \multicolumn{4}{c|}{Outcome equation} \\

   & TMR & Size & Sens. & Spec. & TMR & Size & Sens. & Spec. \\
\hline
 & \multicolumn{8}{c|}{$p=10$} \\
\hline
Normal & 0.885 & 3.129 & 1.000 & 0.981 & 0.932 & 3.074 & 1.000 & 0.989\\
Laplace & 0.815 & 3.224 & 1.000 & 0.968 & 0.874 & 3.142 & 1.000 & 0.980\\
Stepwise$^{*}$ & 0.925 & 3.064 & 0.998 & 0.990 & 0.942 & 3.059 & 1.000 & 0.991\\
ALASSO$^{*}$ & 0.877 & 3.123 & 0.999 & 0.982 & 0.974 & 2.980 & 0.992 & 1.000\\
\hline
 & \multicolumn{8}{c|}{$p=25$} \\
\hline
Normal & 0.911 & 3.091 & 0.999 & 0.996 & 0.935 & 3.056 & 0.998 & 0.997\\
Laplace & 0.891 & 3.121 & 0.999 & 0.994 & 0.898 & 3.100 & 0.999 & 0.995\\
Stepwise$^{*}$ & 0.837 & 3.182 & 1.000 & 0.992 & 0.833 & 3.177 & 1.000 & 0.992\\
ALASSO$^{*}$ & 0.650 & 3.493 & 1.000 & 0.978 & 0.938 & 3.033 & 0.995 & 0.998\\
\hline
 & \multicolumn{8}{c|}{$p=50$} \\
\hline
Normal & 0.904 & 3.086 & 0.998 & 0.998 & 0.945 & 3.042 & 0.997 & 0.999\\
Laplace & 0.880 & 3.125 & 0.999 & 0.997 & 0.922 & 3.070 & 0.997 & 0.998\\
Stepwise$^{*}$ & 0.667 & 3.405 & 1.000 & 0.991 & 0.698 & 3.384 & 1.000 & 0.992\\
ALASSO$^{*}$ & 0.476 & 3.833 & 0.999 & 0.982 & 0.938 & 3.009 & 0.991 & 0.999\\

\hline
\end{tabular}
\caption{Evaluation metrics for $n=1000, \rho=0.5$. ``TMR'' refers to the proportion of seeds where the true model was selected and ``Size'' to the number of variables in the model. Stepwise uses forward selection and ALASSO refers to Adaptive LASSO.\\
$^{*}$ Performance measures for Adaptive LASSO and stepwise selection were only evaluated over replicates that converged and had finite variance.}
\label{table:sim2}
\end{table}
\newpage
\section{Additional tables for different correlations}\label{sec:corrs}

\subsection{$n=500$}

\begin{table}[ht]
\centering
\begin{tabular}{|c|cccc|cccc|}
\hline
 
  \multirow{2}{4em}{Method} & \multicolumn{4}{c|}{Selection equation} & \multicolumn{4}{c|}{Outcome equation} \\

   & TMR & Size & Sens. & Spec. & TMR & Size & Sens. & Spec. \\
\hline
 & \multicolumn{8}{c|}{$p=10$} \\
\hline
Normal & 0.804 & 3.156 & 0.987 & 0.972 & 0.783 & 3.031 & 0.965 & 0.980\\
Laplace & 0.660 & 3.412 & 0.992 & 0.938 & 0.719 & 3.222 & 0.980 & 0.960\\
Stepwise$^{*}$ & 0.855 & 3.007 & 0.975 & 0.988 & 0.808 & 2.938 & 0.955 & 0.990\\
ALASSO$^{*}$ & 0.561 & 3.535 & 0.991 & 0.920 & 0.753 & 2.941 & 0.947 & 0.986\\
\hline
 & \multicolumn{8}{c|}{$p=25$} \\
\hline
Normal & 0.783 & 3.046 & 0.967 & 0.993 & 0.760 & 2.927 & 0.945 & 0.996\\
Laplace & 0.757 & 3.136 & 0.973 & 0.990 & 0.745 & 3.04 & 0.961 & 0.993\\
Stepwise$^{*}$ & 0.691 & 3.192 & 0.970 & 0.987 & 0.690 & 3.154 & 0.966 & 0.988\\
ALASSO$^{*}$ & 0.388 & 4.010 & 0.986 & 0.952 & 0.691 & 2.890 & 0.928 & 0.995\\
\hline
 & \multicolumn{8}{c|}{$p=50$} \\
\hline
Normal & 0.725 & 3.020 & 0.952 & 0.997 & 0.680 & 2.843 & 0.917 & 0.998\\
Laplace & 0.699 & 3.092 & 0.958 & 0.995 & 0.700 & 2.927 & 0.933 & 0.997\\
Stepwise$^{*}$ & 0.495 & 3.562 & 0.972 & 0.986 & 0.520 & 3.450 & 0.965 & 0.988\\
ALASSO$^{*}$ & 0.242 & 4.704 & 0.982 & 0.963 & 0.643 & 2.865 & 0.912 & 0.997\\
\hline
\end{tabular}
\caption{Evaluation metrics for $n=500, \rho=0$. ``TMR'' refers to the proportion of seeds where the true model was selected and ``Size'' to the number of variables in the model. Stepwise uses forward selection and ALASSO refers to Adaptive LASSO.\\
$^{*}$ Performance measures for Adaptive LASSO and stepwise selection were only evaluated over replicates that converged and had finite variance.}
\end{table}


\begin{table}[ht]
\centering
\begin{tabular}{|c|cccc|cccc|}
\hline
 
  \multirow{2}{4em}{Method} & \multicolumn{4}{c|}{Selection equation} & \multicolumn{4}{c|}{Outcome equation} \\

   & TMR & Size & Sens. & Spec. & TMR & Size & Sens. & Spec. \\
\hline
 & \multicolumn{8}{c|}{$p=10$} \\
\hline
Normal & 0.771 & 3.203 & 0.987 & 0.966 & 0.792 & 3.029 & 0.967 & 0.982\\
Laplace & 0.658 & 3.439 & 0.992 & 0.934 & 0.700 & 3.237 & 0.978 & 0.957\\
Stepwise$^{*}$ & 0.820 & 3.003 & 0.968 & 0.986 & 0.809 & 2.962 & 0.960 & 0.988\\ 
ALASSO$^{*}$ & 0.554 & 3.509 & 0.984 & 0.920 & 0.764 & 2.941 & 0.948 & 0.986\\
\hline
 & \multicolumn{8}{c|}{$p=25$} \\
\hline
Normal & 0.754 & 3.053 & 0.964 & 0.993 & 0.766 & 2.946 & 0.948 & 0.995\\
Laplace & 0.724 & 3.153 & 0.972 & 0.989 & 0.749 & 3.034 & 0.957 & 0.993\\
Stepwise$^{*}$ & 0.683 & 3.218 & 0.973 & 0.986 & 0.694 & 3.182 & 0.969 & 0.987\\
ALASSO$^{*}$ & 0.404 & 3.980 & 0.983 & 0.953 & 0.690 & 2.880 & 0.924 & 0.995\\
\hline
 & \multicolumn{8}{c|}{$p=50$} \\
\hline
Normal & 0.716 & 2.965 & 0.941 & 0.997 & 0.677 & 2.804 & 0.910 & 0.998\\
Laplace & 0.706 & 3.038 & 0.951 & 0.996 & 0.691 & 2.897 & 0.926 & 0.997\\
Stepwise$^{*}$ & 0.506 & 3.525 & 0.970 & 0.987 & 0.532 & 3.433 & 0.965 & 0.989\\ 
ALASSO$^{*}$ & 0.264 & 4.609 & 0.976 & 0.964 & 0.567 & 2.750 & 0.881 & 0.998\\
\hline
\end{tabular}
\caption{Evaluation metrics for $n=500, \rho=0.3$. ``TMR'' refers to the proportion of seeds where the true model was selected and ``Size'' to the number of variables in the model. Stepwise uses forward selection and ALASSO refers to Adaptive LASSO.\\
$^{*}$ Performance measures for Adaptive LASSO and stepwise selection were only evaluated over replicates that converged and had finite variance.}
\end{table}


\begin{table}[ht]
\centering
\begin{tabular}{|c|cccc|cccc|}
\hline
 
  \multirow{2}{4em}{Method} & \multicolumn{4}{c|}{Selection equation} & \multicolumn{4}{c|}{Outcome equation} \\

   & TMR & Size & Sens. & Spec. & TMR & Size & Sens. & Spec. \\
\hline
 & \multicolumn{8}{c|}{$p=10$} \\
\hline
Normal & 0.786 & 3.169 & 0.985 & 0.969 & 0.828 & 3.036 & 0.975 & 0.984\\
Laplace & 0.699 & 3.369 & 0.991 & 0.943 & 0.761 & 3.190 & 0.984 & 0.966\\
Stepwise$^{*}$ & 0.813 & 2.969 & 0.961 & 0.988 & 0.804 & 2.945 & 0.956 & 0.989\\
ALASSO$^{*}$ & 0.612 & 3.343 & 0.977 & 0.941 & 0.779 & 2.905 & 0.946 & 0.991\\
\hline
 & \multicolumn{8}{c|}{$p=25$} \\
\hline
Normal & 0.750 & 3.027 & 0.959 & 0.993 & 0.782 & 2.939 & 0.952 & 0.996 \\
Laplace & 0.730 & 3.115 & 0.967 & 0.990 & 0.766 & 3.019 & 0.962 & 0.994\\
Stepwise$^{*}$ & 0.668 & 3.199 & 0.965 & 0.986 & 0.698 & 3.166 & 0.967 & 0.988\\
ALASSO$^{*}$ & 0.426 & 3.788 & 0.973 & 0.961 & 0.713 & 2.895 & 0.932 & 0.996\\
\hline
 & \multicolumn{8}{c|}{$p=50$} \\
\hline
Normal & 0.726 & 2.938 & 0.94 & 0.997 & 0.705 & 2.816 & 0.919 & 0.999 \\
Laplace & 0.711 & 3.002 & 0.947 & 0.997 & 0.703 & 2.898 & 0.930 & 0.998\\
Stepwise$^{*}$ & 0.500 & 3.503 & 0.966 & 0.987 & 0.512 & 3.410 & 0.959 & 0.989\\
ALASSO$^{*}$ & 0.223 & 4.689 & 0.965 & 0.962 & 0.515 & 2.667 & 0.859 & 0.998\\
\hline
\end{tabular}
\caption{Evaluation metrics for $n=500, \rho=0.7$. ``TMR'' refers to the proportion of seeds where the true model was selected and ``Size'' to the number of variables in the model. Stepwise uses forward selection and ALASSO refers to Adaptive LASSO.\\
$^{*}$ Performance measures for Adaptive LASSO and stepwise selection were only evaluated over replicates that converged and had finite variance.}
\end{table}

\clearpage
\subsection{$n=1000$}

\begin{table}[ht]
\centering
\begin{tabular}{|c|cccc|cccc|}
\hline
 
  \multirow{2}{4em}{Method} & \multicolumn{4}{c|}{Selection equation} & \multicolumn{4}{c|}{Outcome equation} \\

   & TMR & Size & Sens. & Spec. & TMR & Size & Sens. & Spec. \\
\hline
 & \multicolumn{8}{c|}{$p=10$} \\
\hline
Normal & 0.895 & 3.112 & 0.999 & 0.984 & 0.927 & 3.079 & 1.000 & 0.989\\
Laplace & 0.833 & 3.189 & 1.000 & 0.973 & 0.870 & 3.149 & 1.000 & 0.979 \\
Stepwise$^{*}$ & 0.932 & 3.049 & 0.997 & 0.992 & 0.948 & 3.054 & 1.000 & 0.992\\
ALASSO$^{*}$ & 0.876 & 3.122 & 0.999 & 0.982 & 0.959 & 2.971 & 0.988 & 0.999\\
\hline
 & \multicolumn{8}{c|}{$p=25$} \\
\hline
Normal & 0.908 & 3.093 & 0.998 & 0.996 & 0.936 & 3.043 & 0.996 & 0.997\\
Laplace & 0.857 & 3.160 & 0.999 & 0.993 & 0.909 & 3.084 & 0.998 & 0.996\\
Stepwise$^{*}$ & 0.825 & 3.192 & 0.999 & 0.991 & 0.830 & 3.177 & 0.998 & 0.992\\
ALASSO$^{*}$ & 0.582 & 3.580 & 0.999 & 0.974 & 0.927 & 3.026 & 0.992 & 0.998\\
\hline
 & \multicolumn{8}{c|}{$p=50$} \\
\hline
Normal & 0.919 & 3.071 & 0.998 & 0.998 & 0.933 & 3.049 & 0.996 & 0.999\\
Laplace & 0.891 & 3.105 & 0.998 & 0.998 & 0.896 & 3.096 & 0.997 & 0.998\\
Stepwise$^{*}$ & 0.694 & 3.354 & 1.000 & 0.992 & 0.680 & 3.382 & 0.999 & 0.992\\
ALASSO$^{*}$ & 0.443 & 3.937 & 1.000 & 0.980 & 0.930 & 3.028 & 0.992 & 0.999\\
\hline
\end{tabular}
\caption{Evaluation metrics for $n=1000, \rho=0$. ``TMR'' refers to the proportion of seeds where the true model was selected and ``Size'' to the number of variables in the model. Stepwise uses forward selection and ALASSO refers to Adaptive LASSO.\\
$^{*}$ Performance measures for Adaptive LASSO and stepwise selection were only evaluated over replicates that converged and had finite variance.}
\end{table}


\begin{table}[ht]
\centering
\begin{tabular}{|c|cccc|cccc|}
\hline
 
  \multirow{2}{4em}{Method} & \multicolumn{4}{c|}{Selection equation} & \multicolumn{4}{c|}{Outcome equation} \\

   & TMR & Size & Sens. & Spec. & TMR & Size & Sens. & Spec. \\
\hline
 & \multicolumn{8}{c|}{$p=10$} \\
\hline
Normal & 0.880 & 3.121 & 0.999 & 0.982 & 0.928 & 3.074 & 0.999 & 0.989\\
Laplace & 0.812 & 3.209 & 0.999 & 0.970 & 0.878 & 3.135 & 1.000 & 0.981\\
Stepwise$^{*}$ & 0.931 & 3.056 & 0.998 & 0.991 & 0.942 & 3.055 & 0.999 & 0.992\\
ALASSO$^{*}$ & 0.853 & 3.146 & 0.998 & 0.978 & 0.973 & 2.983 & 0.993 & 0.999\\
\hline
 & \multicolumn{8}{c|}{$p=25$} \\
\hline
Normal & 0.917 & 3.082 & 0.999 & 0.996 & 0.919 & 3.066 & 0.997 & 0.997\\
Laplace & 0.886 & 3.125 & 1.000 & 0.994 & 0.892 & 3.107 & 0.999 & 0.995\\
Stepwise$^{*}$ 0.828 & 3.183 & 1.000 & 0.992 & 0.825 & 3.178 & 0.999 \\
ALASSO$^{*}$ & 0.619 & 3.519 & 1.000 & 0.976 & 0.929 & 3.035 & 0.993 & 0.998\\
\hline
 & \multicolumn{8}{c|}{$p=50$} \\
\hline
Normal & 0.903 & 3.088 & 0.998 & 0.998 & 0.939 & 3.036 & 0.995 & 0.999\\
Laplace & 0.873 & 3.123 & 0.998 & 0.997 & 0.923 & 3.060 & 0.996 & 0.998\\
Stepwise$^{*}$ & 0.664 & 3.398 & 0.999 & 0.991 & 0.679 & 3.383 & 0.999 & 0.992\\
ALASSO$^{*}$ & 0.458 & 3.906 & 1.000 & 0.981 & 0.940 & 3.017 & 0.992 & 0.999\\
\hline
\end{tabular}
\caption{Evaluation metrics for $n=1000, \rho=0.3$. ``TMR'' refers to the proportion of seeds where the true model was selected and ``Size'' to the number of variables in the model. Stepwise uses forward selection and ALASSO refers to Adaptive LASSO.\\
$^{*}$ Performance measures for Adaptive LASSO and stepwise selection were only evaluated over replicates that converged and had finite variance.}
\end{table}


\begin{table}[ht]
\centering
\begin{tabular}{|c|cccc|cccc|}
\hline
 
  \multirow{2}{4em}{Method} & \multicolumn{4}{c|}{Selection equation} & \multicolumn{4}{c|}{Outcome equation} \\

   & TMR & Size & Sens. & Spec. & TMR & Size & Sens. & Spec. \\
\hline
 & \multicolumn{8}{c|}{$p=10$} \\
\hline
Normal & 0.899 & 3.116 & 1.000 & 0.983 & 0.938 & 3.064 & 1.000 & 0.991\\
Laplace & 0.839 & 3.201 & 1.000 & 0.971 & 0.878 & 3.133 & 1.000 & 0.981 \\
Stepwise$^{*}$ & 0.939 & 3.059 & 0.999 & 0.991 & 0.940 & 3.057 & 0.999 & 0.992\\
ALASSO$^{*}$ & 0.915 & 3.080 & 0.998 & 0.988 & 0.984 & 2.986 & 0.995 & 1.000\\
\hline
 & \multicolumn{8}{c|}{$p=25$} \\
\hline
Normal & 0.921 & 3.079 & 0.999 & 0.996 & 0.943 & 3.048 & 0.998 & 0.998\\
Laplace & 0.877 & 3.132 & 1.000 & 0.994 & 0.914 & 3.084 & 0.999 & 0.996\\
Stepwise$^{*}$ & 0.830 & 3.182 & 1.000 & 0.992 & 0.827 & 3.188 & 1.000 & 0.991\\
ALASSO$^{*}$ & 0.667 & 3.449 & 0.999 & 0.980 & 0.945 & 3.035 & 0.997 & 0.998\\
\hline
 & \multicolumn{8}{c|}{$p=50$} \\
\hline
Normal & 0.911 & 3.089 & 0.999 & 0.998 & 0.940 & 3.048 & 0.997 & 0.999 \\
Laplace & 0.882 & 3.126 & 1.000 & 0.997 & 0.924 & 3.075 & 0.998 & 0.998\\
Stepwise$^{*}$ & 0.668 & 3.399 & 1.000 & 0.992 & 0.688 & 3.387 & 1.000 & 0.992\\
ALASSO$^{*}$ & 0.477 & 3.847 & 0.999 & 0.982 & 0.934 & 3.029 & 0.993 & 0.999\\
\hline
\end{tabular}
\caption{Evaluation metrics for $n=1000, \rho=0.7$. ``TMR'' refers to the proportion of seeds where the true model was selected and ``Size'' to the number of variables in the model. Stepwise uses forward selection and ALASSO refers to Adaptive LASSO.\\
$^{*}$ Performance measures for Adaptive LASSO and stepwise selection were only evaluated over replicates that converged and had finite variance.}
\end{table}


\clearpage
\section{Class II prior comparison}\label{sec:class2}

Recall that $\tilde{\sigma}^{2} = \sigma^{2}(1-\rho^{2})$. For $\rho = 0$, the Class II prior is exactly dependent on the outcome variance, and the prior is shrunk for $\lvert \rho \rvert > 0$. In the simulation studies, we chose $\sigma^{2} = 1$, so the Class II prior performs similarly to the Class I prior for small $\rho$; for this reason, we present results for the most extreme case, $\rho = 0.7$.

\begin{table}[ht]
\centering
\begin{tabular}{|c|cccc|cccc|}
\hline
 
  \multirow{2}{4em}{Method} & \multicolumn{4}{c|}{Selection equation} & \multicolumn{4}{c|}{Outcome equation} \\

   & TMR & Size & Sens. & Spec. & TMR & Size & Sens. & Spec. \\
\hline
 & \multicolumn{8}{c|}{$p=10$} \\
\hline
Class I normal & 0.786 & 3.169 & 0.985 & 0.969 & 0.828 & 3.036 & 0.975 & 0.984\\
Class II normal & 0.787 & 3.191 & 0.987 & 0.967 & 0.790 & 3.122 & 0.980 & 0.974\\
Class I Laplace & 0.699 & 3.369 & 0.991 & 0.943 & 0.761 & 3.190 & 0.984 & 0.966\\
Class II Laplace & 0.674 & 3.45 & 0.993 & 0.933 & 0.676 & 3.362 & 0.987 & 0.943\\
\hline
 & \multicolumn{8}{c|}{$p=25$} \\
\hline
Class I normal & 0.750 & 3.027 & 0.959 & 0.993 & 0.782 & 2.939 & 0.952 & 0.996 \\
Class II normal & 0.760 & 3.044 & 0.963 & 0.993 & 0.774 & 2.984 & 0.957 & 0.995\\
Class I Laplace & 0.730 & 3.115 & 0.967 & 0.990 & 0.766 & 3.019 & 0.962 & 0.994\\
Class II Laplace & 0.732 & 3.128 & 0.969 & 0.990 & 0.753 & 3.073 & 0.967 & 0.992\\
\hline
 & \multicolumn{8}{c|}{$p=50$} \\
\hline
Class I normal & 0.726 & 2.938 & 0.940 & 0.997 & 0.705 & 2.816 & 0.919 & 0.999 \\
Class II normal & 0.745 & 2.952 & 0.947 & 0.998 & 0.705 & 2.858 & 0.924 & 0.998\\
Class I Laplace & 0.711 & 3.002 & 0.947 & 0.997 & 0.703 & 2.898 & 0.930 & 0.998\\
Class II Laplace & 0.717 & 3.035 & 0.953 & 0.996 & 0.719 & 2.915 & 0.935 & 0.998\\
\hline
\end{tabular}
\caption{Comparison between Class I and Class II priors for $n=500, \rho=0.7$. ``TMR'' refers to the proportion of seeds where the true model was selected and ``Size'' to the number of variables in the model.}
\end{table}

\begin{table}[ht]
\centering
\begin{tabular}{|c|cccc|cccc|}
\hline
 
  \multirow{2}{4em}{Method} & \multicolumn{4}{c|}{Selection equation} & \multicolumn{4}{c|}{Outcome equation} \\

   & TMR & Size & Sens. & Spec. & TMR & Size & Sens. & Spec. \\
\hline
 & \multicolumn{8}{c|}{$p=10$} \\
\hline
Class I normal & 0.899 & 3.116 & 1.000 & 0.983 & 0.938 & 3.064 & 1.000 & 0.991\\
Class II normal & 0.893 & 3.125 & 1.000 & 0.982 & 0.902 & 3.104 & 1.000 & 0.985\\
Class I Laplace & 0.839 & 3.201 & 1.000 & 0.971 & 0.878 & 3.133 & 1.000 & 0.981 \\
Class II Laplace & 0.813 & 3.235 & 1.000 & 0.966 & 0.813 & 3.224 & 1.000 & 0.968\\
\hline
 & \multicolumn{8}{c|}{$p=25$} \\
\hline
Class I normal & 0.921 & 3.079 & 0.999 & 0.996 & 0.943 & 3.048 & 0.998 & 0.998\\
Class II normal & 0.915 & 3.086 & 0.999 & 0.996 & 0.922 & 3.068 & 0.998 & 0.997\\
Class I Laplace & 0.877 & 3.132 & 1.000 & 0.994 & 0.914 & 3.084 & 0.999 & 0.996\\
Class II Laplace & 0.867 & 3.15 & 1.000 & 0.993 & 0.885 & 3.116 & 0.999 & 0.995\\
\hline
 & \multicolumn{8}{c|}{$p=50$} \\
\hline
Class I normal & 0.911 & 3.089 & 0.999 & 0.998 & 0.940 & 3.048 & 0.997 & 0.999 \\
Class II normal & 0.921 & 3.081 & 1.000 & 0.998 & 0.920 & 3.078 & 0.998 & 0.998\\
Class I Laplace & 0.882 & 3.126 & 1.000 & 0.997 & 0.924 & 3.075 & 0.998 & 0.998\\
Class II Laplace & 0.872 & 3.141 & 1.000 & 0.997 & 0.903 & 3.102 & 0.999 & 0.998\\
\hline
\end{tabular}
\caption{Comparison between Class I and Class II priors for $n=1000, \rho=0.7$. ``TMR'' refers to the proportion of seeds where the true model was selected  and ``Size'' to the number of variables in the model.}
\end{table}

The two perform fairly similarly, with Class II priors including more variables on average than Class I priors (reflected in the model size).

\clearpage

\section{MCMC diagnostics}\label{sec:mcmc}
To assess convergence of the Gibbs sampler, we look at the Effective Sample Size (ESS) of parameter estimates for different replicates in the Simulation study. For each of the $1,000$ replicates in the main simulation study scenarios (Section \ref{sec:sim_main}, i.e. $\rho = 0.5$), we recorded the ESS of each variable, with the total chain length being $8,750$. For each variable, we pooled results for $p=10, 25$ and $50$ as we found that the ESS did not differ significantly with number of variables. For simplicity of presentation, we pool the active variables in each equation together, and inactive variables in each equation together, alongside presenting the ESS of $\rho$.

\begin{figure}[h]
    \centering
    \includegraphics[width=0.7\linewidth]{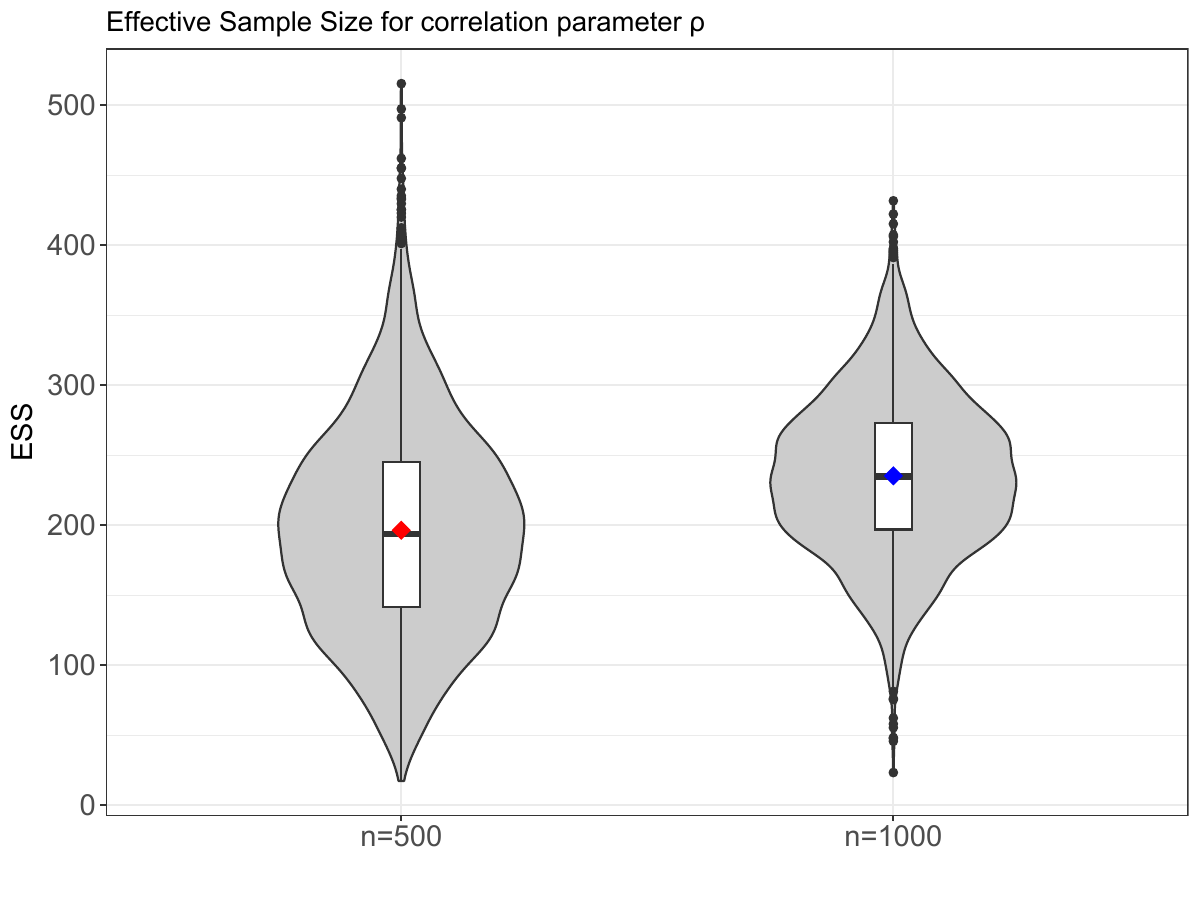}
    \caption{Violin plot of effective sample sizes for $\rho$ across simulation replicates.}
    \label{fig:essrho}
\end{figure}

Figure \ref{fig:essrho} shows the ESS of $\rho$ for $n=500$ and $n=1000$. The median ESS is just below $200$ for $n=500$, which is small relative to the chain length of $8,750$. This suggests that the mixing is very slow. It should be noted that $\rho$ is a particularly difficult variable to estimate in sample selection models: it typically has a large standard error (such as in the ambulatory data: see Table \ref{table:amb}) while also having issues with convergence of maximum likelihood estimation (see \ref{sec:convergence}). For instance, in one case where $\rho$ has an ESS of just $23$, the cause is severe bimodality in the true model from the generated data. Regardless, the Gibbs sampler still mixes slowly, even for $n=1,000$ (where the median ESS is still below $250$).

\begin{figure}[h]
    \centering
    \includegraphics[width=0.7\linewidth]{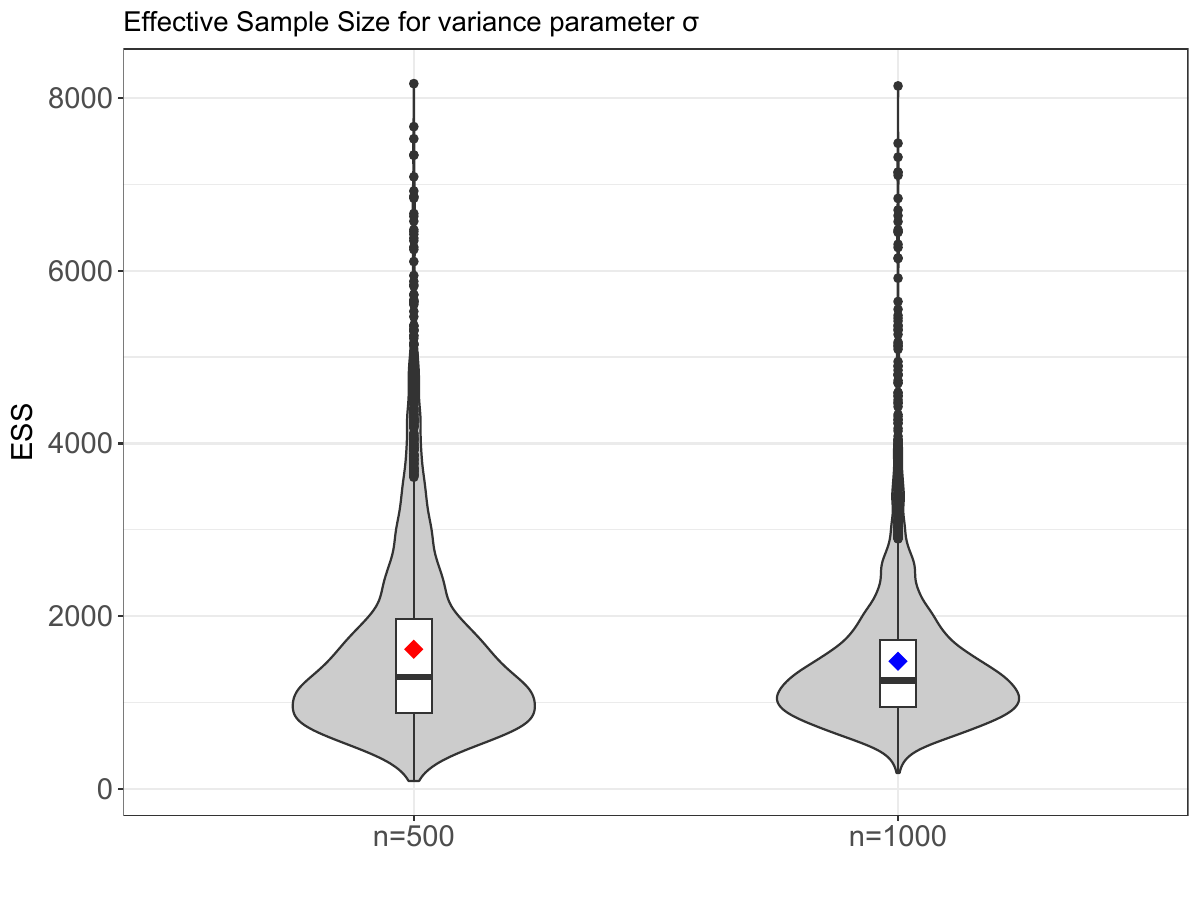}
    \caption{Violin plot for effective sample sizes of $\sigma$ across simulation replicates.}
    \label{fig:esssigma}
\end{figure}

It should also be noted that $\rho$ and $\sigma$ are not sampled directly, but instead $\tilde{\rho} =\rho\sigma$ and $\tilde{\sigma} = \sqrt{\sigma^{2}(1-\rho^{2})}$. Figure \ref{fig:esssigma} shows that the effective sample sizes for $\sigma$ are significantly larger than those for $\rho$, but internally (not shown here) we found that the effective sample sizes for $\tilde{\rho}$ and $\tilde{\sigma}$ were closer to that of $\rho$ than $\sigma$. This suggests that the problematic part of the error distribution to model is not the variance but the correlation. Additionally, the ratio of effective sample size to chain length is consistent with those reported by \cite{wiemann2022correcting}, suggesting that the problem may be inherent to $\rho$.

\begin{figure}[h]
    \centering
    \includegraphics[width=0.7\linewidth]{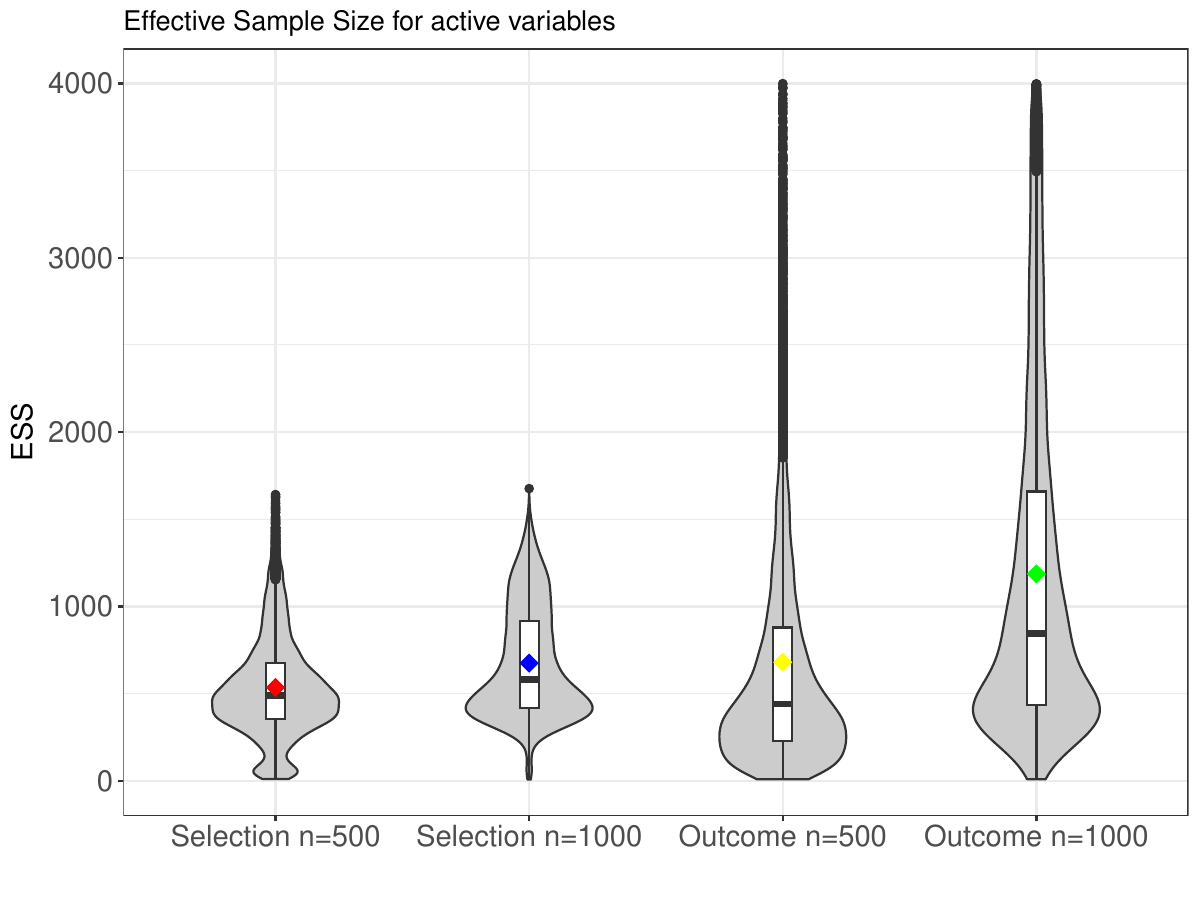}
    \caption{Violin plot for effective sample sizes of active variables across simulation replicates, with each equation pooled separately. The long tail for $n=1000$ outside the graph has been cut off for clarity.}
    \label{fig:essactive}
\end{figure}

\begin{figure}[h]
    \centering
    \includegraphics[width=0.7\linewidth]{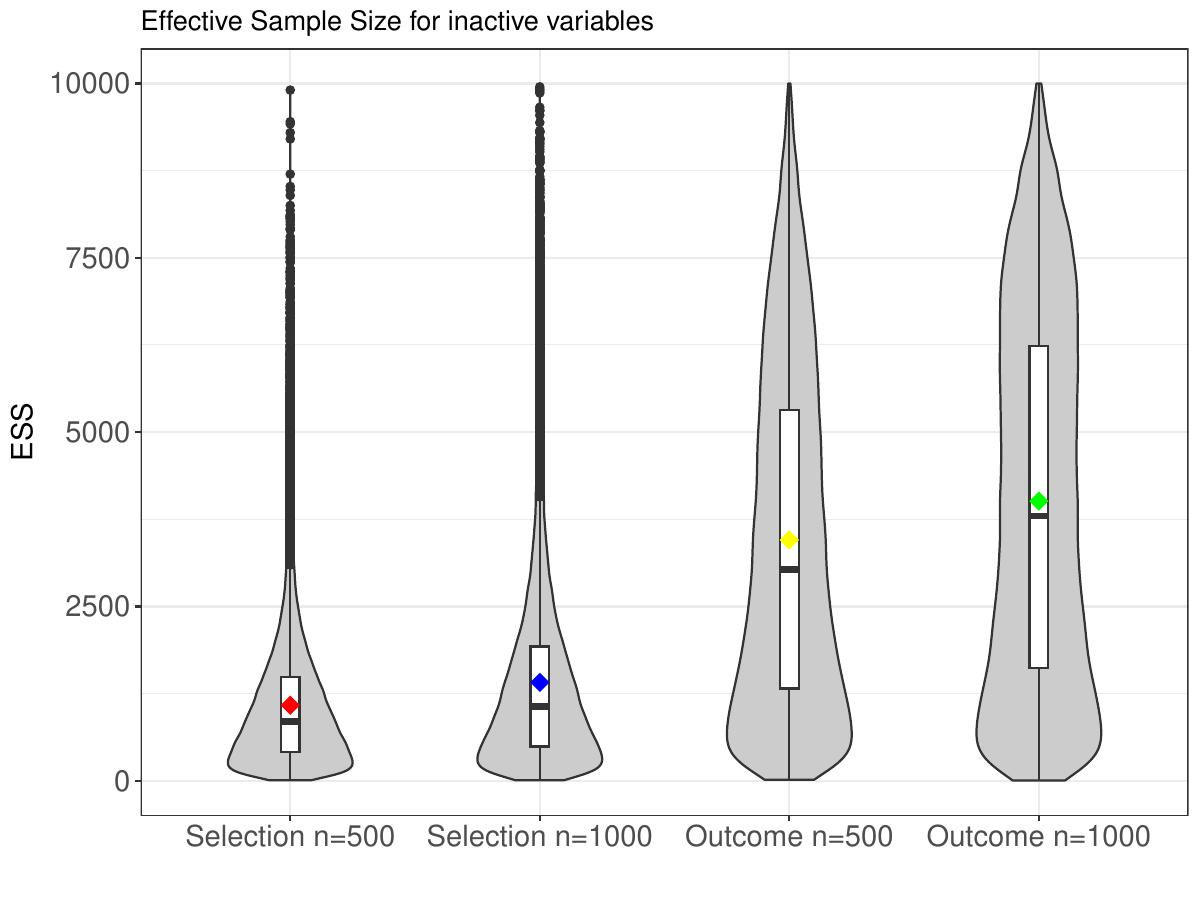}
    \caption{Violin plot for effective sample sizes of inactive variables across simulation replicates, with each equation pooled separately. The long tail for $n=1000$ outside the graph has been cut off for clarity.}
    \label{fig:essinactive}
\end{figure}

Figures \ref{fig:essactive} and \ref{fig:essinactive} show the ESS values for active and inactive variables respectively in each equation. In each case, the graph is cut off for clarity (there is a long but narrow tail in each $n=1000$ case). Unsurprisingly, the mixing in the selection equation is on average slower than the mixing in the outcome equation, while the median ESS varies from around $500$ to almost $4,000$, with inactive variables having higher ESS than active ones.

Aside from the active variables in the selection equation, there is a not-insignificant density towards very low ESS values. There are two causes of this.
Firstly, as $\rho$ changes values, so will the distribution of the other parameters conditional on $\rho$. Because $\rho$ mixes slowly, it will slow down the mixing for other active parameters too. This is particularly true for the intercept and large effects in the outcome equation: $\rho\sigma$ is sampled as a ``covariate'' alongside the outcome parameters in the Gibbs sampler in Section \ref{sec:vansampler}, so any large effects have highly correlated parameter samples. As such, a small effective sample size for $\rho$ will lead to small effective sample sizes for other correlated parameters.
Secondly, for variables that have posterior inclusion probability not close to $0$ or $1$, the posterior will be bimodal (though the mode corresponding to the slab can be very small if the posterior inclusion probability is small, e.g. around $0.2$). This is exacerbated by the correlation of the parameter with $\rho$ when it is included in the model, leading to very slow mixing around one of the modes.

It should also be noted that there are cases where the effective sample size exceeds the chain length ($8,750$) for spurious variables. Gibbs samplers cannot be theoretically antithetic, but sometimes finite sample autocorrelation can be negative. For some spurious variables with very low inclusion probability, this can occur by chance.

For the simulation study, the performance measures were taken on the median model. In particular, the median model only differs if a posterior inclusion probability changes from one side of $0.5$ to the other. We found that in practice, that posterior inclusion probabilities did not drastically differ even when effective sample sizes were small and mixing was slow (potentially because the areas of highest posterior density are those being sampled frequently) so that $10,000$ iterations was sufficient to stabilize the median model. We prioritized additional Monte Carlo replicates to reduce variation caused by different data samples, as opposed to variance caused by chains. To verify this, we performed the simulation study in Section \ref{sec:sim_main} for $n=500$, $p=10$ only, with a chain length of $50,000$ instead of $10,000$, and a thinning interval of $5$ for memory purposes. We found that the results differed very marginally between the chain lengths.

Regardless, for practical applications we would suggest running the chain for longer, e.g. $n=50,000$ as in Section \ref{sec:applications}. With the computational times reported in Appendix \ref{sec:comptimes} for $10,000$ iterations, this is not infeasible and should allow for an effective sample size of all parameters above $500$. Furthermore, diagnosing convergence of the chain can be done post-hoc: more iterations can easily be added if necessary.

\begin{table}[ht]
\centering
\begin{tabular}{|c|cccc|cccc|}
\hline
 
  \multirow{2}{4em}{Method} & \multicolumn{4}{c|}{Selection equation} & \multicolumn{4}{c|}{Outcome equation} \\

   & TMR & Size & Sens. & Spec. & TMR & Size & Sens. & Spec. \\
\hline
Normal $10,000$ & 0.784 & 3.162 & 0.984 & 0.970 & 0.797 & 3.045 & 0.971 & 0.981\\
Normal $50,000$ & 0.783 & 3.173 & 0.985 & 0.969 & 0.794 & 3.045 & 0.970 & 0.981\\
Laplace $10,000$ & 0.669 & 3.402 & 0.991 & 0.939 & 0.730 & 3.219 & 0.982 & 0.960\\
Laplace $50,000$ & 0.666 & 3.402 & 0.991 & 0.939 & 0.732 & 3.215 & 0.982 & 0.962\\
\hline
\end{tabular}
\caption{Comparison of different chain lengths for $n=500$, $p=10$, $\rho = 0.5$ for spike-and-slab priors. For chain length $50,000$, a thinning interval of $5$ was used.}
\label{table:longchain}
\end{table}

Finally, we test the impact of different initial values by running a small simulation study for $n=500, p=10, \rho=0.5$, we simulate the data as before, but run $10$ chains (each for $10,000$ and discarding $1,250$ as burn-in) from different initial values. We record the point-estimate of the Gelman-Rubin statistic $\widehat{R}$ taken for these $10$ chains. We repeat this for $100$ replicates and summarize the quantiles of the sampled $\widehat{R}$ below.

To choose random initial values, we chose an initial model at complete random and then sampled initial values for active coefficients from normal distributions (appropriately transforming $\sigma$ and $\rho$ to be in $(0,\infty)$ and $(-1,1)$ respectively). It should be noted that this is an extreme scenario as the initial model may not only include many irrelevant variables but also may exclude variables with very large effect sizes and coefficients may have completely opposite signs to their true values, as opposed to initializing around the MLE which is more likely to start in a region of high posterior density.

Table \ref{table:rhat} shows that the $\widehat{R}$ is very close to $1$ for almost all simulation replicates, which suggests the $10$ chains converged to the same distribution for almost all replicates. The only variable that displays a departure from this is in the variable $\rho$, where a single replicate has its $\widehat{R}$ exceed $1.05$. Given the potentially extreme initial values used and the lower value of $\widehat{R}$ for $\rho$ in all other replicates, we do not think this is problematic, but this can be remedied by running the chain for, say, $50,000$ replicates, which we suggest for all practical applications regardless.

\begin{table}[ht]
\centering
\begin{tabular}{|c|ccccccc|}
\hline
\multirow{2}{4em}{Variables} & \multicolumn{7}{c|}{$\widehat{R}$ point-estimate quantile} \\
 & 0\% & 2.5\% & 25\% & 50\% & 75\% & 97.5\% & 100\% \\
 \hline
 Active $\bm{\alpha}$ & 1.000 & 1.001 & 1.002 & 1.002 & 1.004 & 1.009 & 1.012 \\
 Inactive $\bm{\alpha}$ & 1.000 & 1.001 & 1.002 & 1.003 & 1.004 & 1.008 & 1.020 \\
 Active $\bm{\beta}$ & 1.000 & 1.000 & 1.001 & 1.002 & 1.004 & 1.012 & 1.025 \\
 Inactive $\bm{\beta}$ & 1.000 & 1.000 & 1.000 & 1.001 & 1.001 & 1.003 & 1.005 \\
$\rho$ & 1.002 & 1.002 & 1.004 & 1.006 & 1.010 & 1.024 & 1.053 \\
$\sigma$ & 1.000 & 1.000 & 1.001 & 1.001 & 1.001 & 1.003 & 1.005 \\
\hline
\end{tabular}
\caption{Table testing the impact of different initial values on point estimators of $\widehat{R}$. The point-estimates were recorded for $100$ different replicates, hence the quantiles of the point-estimates being reported. $\bm{\alpha}$ refers to selection equation variables and $\bm{\beta}$ to outcome equation variables.}
\label{table:rhat}
\end{table}
\clearpage

\clearpage
\section{Additional tables for sensitivity analysis}\label{sec:sens_tables}
\subsection{Varying slab variance for ambulatory data}
\begin{table}[ht]
\centering
\begin{tabular}{|c|ccc|ccc|ccc|}
\hline \multirow{2}{*}{Slab variance scaling} & \multicolumn{3}{|c|}{$\times 1$} & \multicolumn{3}{|c|}{$\times 0.25$} & \multicolumn{3}{|c|}{$\times 4$}\\
  & PIP & Est. & S.D.& PIP & Est.& S.D.& PIP & Est. & S.D.\\
 \hline
 & \multicolumn{9}{c|}{Selection equation}\\
 \hline
(Intercept) & - & 1.276 & 0.038 & - & 1.268 & 0.037 & - & 1.275 & 0.038 \\

educ & 1.000 & 0.177 & 0.032 & 1.000 & 0.172 & 0.032 & 0.996 & 0.181 & 0.033 \\

age & 0.949 & 0.112 & 0.039 & 0.977 & 0.109 & 0.034 & 0.914 & 0.115 & 0.044 \\

income & 0.349 & 0.007 & 0.042 & 0.471 & 0.012 & 0.044 & 0.206 & 0.005 & 0.037 \\

female & 1.000 & 0.320 & 0.031 & 1.000 & 0.319 & 0.030 & 1.000 & 0.320 & 0.031 \\

totchr & 1.000 & 0.602 & 0.055 & 1.000 & 0.585 & 0.052 & 1.000 & 0.604 & 0.055 \\

blhisp &  1.000 & -0.171 &  0.029 &  1.000 & -0.169 &  0.029 &  1.000 & -0.174 &  0.029 \\

ins & 0.571 & 0.048 & 0.046 & 0.693 & 0.062 & 0.044 & 0.388 & 0.009 & 0.044 \\
 \hline
 & \multicolumn{9}{c|}{Outcome equation}\\
 \hline
(Intercept) & - & 6.563 & 0.060 & - & 6.562 & 0.057 & - & 6.566 & 0.059 \\

educ & 0.116 & 0.004 & 0.018 & 0.196 & 0.005 & 0.021 & 0.062 & 0.003 & 0.014 \\

age & 1.000 & 0.230 & 0.026 & 1.000 & 0.229 & 0.026 & 1.000 & 0.231 & 0.026 \\

female & 1.000 & 0.158 & 0.032 & 1.000 & 0.158 & 0.030 & 1.000 & 0.157 & 0.031 \\

totchr & 1.000 & 0.399 & 0.031 & 1.000 & 0.398 & 0.030 & 1.000 & 0.398 & 0.031 \\

blhisp &  0.895 & -0.094 &  0.039 &  0.952 & -0.095 &  0.033 &  0.844 & -0.093 &  0.042 \\

ins &  0.033 & -0.001 &  0.008 &  0.079 & -0.002 &  0.010 &  0.017 & -0.001 &  0.008 \\
 \hline
$\sigma$ & - & 1.286 & 0.024 & - & 1.286 & 0.023 & - & 1.287 & 0.024\\
 $\rho$ & - & -0.265 & 0.154 & - & -0.266 &  0.145 & - & -0.273 &  0.150 \\
\hline
\end{tabular}\label{table:amb_slab}
\caption{Results from ambulatory data from varying the slab variances. The first column refers to what $\tau_{1,\alpha}$ and $\tau_{1,\beta}$ have been scaled by. The elicitation is otherwise the same as the previous data studies.}
\end{table}

\subsection{Varying spike variance for ambulatory data}
\begin{table}[ht]
\centering
\begin{tabular}{|c|ccc|ccc|ccc|}
\hline \multirow{2}{*}{Spike variances scaling} & \multicolumn{3}{|c|}{$\times 1$} & \multicolumn{3}{|c|}{$\times 0.25$} & \multicolumn{3}{|c|}{$\times 4$}\\
  & PIP & Est. & S.D.& PIP & Est. & S.D.& PIP & Est. & S.D.\\
 \hline
 & \multicolumn{9}{c|}{Selection equation}\\
 \hline
(Intercept) & - & 1.276 & 0.038 & - & 1.278 & 0.038 & - & 1.275 & 0.039 \\

educ & 1.000 & 0.177 & 0.032 & 1.000 & 0.177 & 0.032 & 1.000 & 0.175 & 0.031 \\

age & 0.949 & 0.112 & 0.039 & 0.944 & 0.110 & 0.040 & 0.910 & 0.110 & 0.041 \\

income & 0.349 & 0.007 & 0.042 & 0.402 & 0.004 & 0.045 & 0.301 & 0.016 & 0.039 \\

female & 1.000 & 0.320 & 0.031 & 1.000 & 0.322 & 0.031 & 1.000 & 0.321 & 0.030 \\

totchr & 1.000 & 0.602 & 0.055 & 1.000 & 0.605 & 0.055 & 1.000 & 0.601 & 0.056 \\

blhisp &  1.000 & -0.171 &  0.029 &  1.000 & -0.171 &  0.029 &  1.000 & -0.172 &  0.029 \\

ins & 0.571 & 0.048 & 0.046 & 0.626 & 0.056 & 0.046 & 0.487 & 0.031 & 0.042 \\
 \hline
 & \multicolumn{9}{c|}{Outcome equation}\\
 \hline
(Intercept) & - & 6.563 & 0.060 & - & 6.556 & 0.059 & - & 6.564 & 0.060 \\

educ & 0.116 & 0.004 & 0.018 & 0.131 & 0.001 & 0.019 & 0.103 & 0.011 & 0.019 \\

age & 1.000 & 0.230 & 0.026 & 1.000 & 0.231 & 0.026 & 1.000 & 0.231 & 0.026 \\

female & 1.000 & 0.158 & 0.032 & 1.000 & 0.161 & 0.031 & 0.999 & 0.158 & 0.032 \\

totchr & 1.000 & 0.399 & 0.031 & 1.000 & 0.401 & 0.031 & 1.000 & 0.398 & 0.032 \\

blhisp &  0.895 & -0.094 &  0.039 &  0.935 & -0.097 &  0.036 &  0.816 & -0.090 & 0.040 \\

ins &  0.033 & -0.001 &  0.008 &  0.036 & 0.000 &  0.006 &  0.039 & -0.004 &  0.013 \\
 \hline
$\sigma$ & - & 1.286 & 0.024 & - & 1.284 & 0.023 & - & 1.286 & 0.024\\
 $\rho$ & - & -0.265 &  0.154 & - & -0.245 & 0.152 & - & -0.268 &  0.155 \\
\hline
\end{tabular}\label{table:amb_spike}
\caption{Results from ambulatory data from varying the spike variances. The first column refers to what $\tau_{0,\alpha}$ and $\tau_{0,\beta}$ have been scaled by. The elicitation is otherwise the same as the previous data studies.}
\end{table}
\clearpage

\subsection{Varying slab for RAND data}
\begin{table}[ht]
\centering
\resizebox{1 \textwidth}{!}{
\begin{tabular}{|c|ccc|ccc|ccc|}
\hline \multirow{2}{*}{Slab variance scaling} & \multicolumn{3}{|c|}{$\times 1$} & \multicolumn{3}{|c|}{$\times 0.25$} & \multicolumn{3}{|c|}{$\times 4$}\\
 &  PIP & Est. & S.D. &  PIP & Est. & S.D. &  PIP & Est. & S.D.\\
 \hline
 & \multicolumn{9}{c|}{Selection equation}\\
 \hline
(Intercept) & - & 0.813 & 0.020 & - & 0.813 & 0.020 & - & 0.813 & 0.020 \\

logc &  1.000 & -0.229 &  0.030 &  1.000 & -0.226 &  0.033 &  1.000 & -0.229 &  0.029 \\

idp &  0.094 & -0.001 &  0.012 &  0.206 & -0.002 &  0.018 &  0.047 & -0.001 &  0.009 \\

lpi & 0.919 & 0.072 & 0.028 & 0.930 & 0.069 & 0.027 & 0.899 & 0.073 & 0.030 \\

fmde & 0.090 & 0.001 & 0.023 & 0.155 & 0.001 & 0.028 & 0.068 & 0.001 & 0.023 \\

physlm & 0.998 & 0.099 & 0.024 & 0.997 & 0.096 & 0.024 & 1.000 & 0.100 & 0.023 \\

disea & 1.000 & 0.154 & 0.023 & 1.000 & 0.152 & 0.023 & 1.000 & 0.155 & 0.023 \\

hlthg & 0.020 & 0.000 & 0.004 & 0.049 & 0.000 & 0.006 & 0.007 & 0.000 & 0.004 \\

hlthf & 0.131 & 0.001 & 0.018 & 0.336 & 0.003 & 0.026 & 0.051 & 0.001 & 0.011 \\

hlthp & 0.844 & 0.077 & 0.038 & 0.960 & 0.084 & 0.030 & 0.724 & 0.069 & 0.042 \\

linc & 0.640 & 0.048 & 0.034 & 0.849 & 0.059 & 0.029 & 0.552 & 0.037 & 0.035 \\

lfam &  0.019 &  0.000 &  0.005 &  0.046 & 0.000 &  0.006 &  0.009 &  0.000 &  0.004 \\

educdec & 0.922 & 0.071 & 0.027 & 0.961 & 0.073 & 0.024 & 0.833 & 0.067 & 0.032 \\

xage &  0.031 & 0.000 &  0.007 &  0.072 & 0.000 &  0.009 &  0.012 &  0.000 &  0.005 \\

female & 1.000 & 0.196 & 0.024 & 1.000 & 0.194 & 0.024 & 1.000 & 0.196 & 0.024 \\

child & 0.057 & 0.001 & 0.011 & 0.082 & 0.001 & 0.012 & 0.016 & 0.001 & 0.006 \\

fchild &  1.000 & -0.143 &  0.024 &  1.000 & -0.141 &  0.025 &  1.000 & -0.143 &  0.023 \\

black &  1.000 & -0.227 &  0.021 &  1.000 & -0.223 &  0.020 &  1.000 & -0.228 &  0.021 \\
 \hline
 & \multicolumn{9}{c|}{Outcome equation}\\
 \hline
(Intercept) & - & 3.551 & 0.039 & - & 3.552 & 0.040 & - & 3.544 & 0.039 \\

logc &  1.000 & -0.229 &  0.027 &  0.994 & -0.226 &  0.036 &  1.000 & -0.230 &  0.026 \\

idp &  0.033 & -0.001 &  0.007 &  0.102 & -0.001 &  0.015 &  0.013 & 0.000 &  0.005 \\

lpi & 0.046 & 0.001 & 0.010 & 0.085 & 0.001 & 0.012 & 0.022 & 0.000 & 0.007 \\

fmde &  0.037 & 0.000 &  0.011 &  0.106 & 0.000 &  0.024 &  0.020 & 0.000 &  0.009 \\

physlm & 1.000 & 0.124 & 0.025 & 1.000 & 0.121 & 0.025 & 1.000 & 0.129 & 0.025 \\

disea & 1.000 & 0.205 & 0.027 & 1.000 & 0.201 & 0.027 & 1.000 & 0.208 & 0.027 \\

hlthg & 0.236 & 0.002 & 0.026 & 0.497 & 0.008 & 0.033 & 0.084 & 0.001 & 0.017 \\

hlthf & 0.740 & 0.066 & 0.040 & 0.898 & 0.082 & 0.037 & 0.507 & 0.021 & 0.040 \\

hlthp & 0.964 & 0.100 & 0.031 & 0.993 & 0.108 & 0.026 & 0.886 & 0.092 & 0.038 \\

linc & 0.997 & 0.138 & 0.034 & 1.000 & 0.147 & 0.031 & 1.000 & 0.130 & 0.034 \\

lfam &  0.853 & -0.079 &  0.037 &  0.916 & -0.080 &  0.033 &  0.715 & -0.072 &  0.043 \\

educdec & 0.076 & 0.001 & 0.014 & 0.175 & 0.001 & 0.021 & 0.036 & 0.001 & 0.010 \\

xage & 0.749 & 0.115 & 0.071 & 0.783 & 0.100 & 0.063 & 0.736 & 0.136 & 0.079 \\

female & 1.000 & 0.279 & 0.035 & 1.000 & 0.271 & 0.034 & 1.000 & 0.285 & 0.035 \\

child &  0.670 & -0.110 &  0.083 &  0.799 & -0.120 &  0.073 &  0.544 & -0.079 &  0.092 \\

fchild &  1.000 & -0.234 &  0.044 &  1.000 & -0.222 &  0.042 &  1.000 & -0.244 &  0.044 \\

black &  1.000 & -0.205 &  0.030 &  1.000 & -0.203 &  0.030 &  1.000 & -0.207 &  0.030 \\
 \hline
$\sigma$ & - & 1.571 & 0.029 & - & 1.568 & 0.030 & - & 1.577 & 0.030\\
 $\rho$ & - & 0.729 & 0.039 & - & 0.725 & 0.041 & - & 0.736 & 0.040 \\
\hline
\end{tabular}}\label{table:rand_slab}
\caption{Results from RAND data from varying the slab variances. The first column refers to what $\tau_{1,\alpha}$ and $\tau_{1,\beta}$ have been scaled by. The elicitation is otherwise the same as the previous data studies.}
\end{table}

\clearpage
\subsection{Varying spike for RAND data}
\begin{table}[ht]
\centering
\resizebox{1 \textwidth}{!}{
\begin{tabular}{|c|ccc|ccc|ccc|}
\hline \multirow{2}{*}{Spike variance scaling} & \multicolumn{3}{|c|}{$\times 1$} & \multicolumn{3}{|c|}{$\times 0.25$} & \multicolumn{3}{|c|}{$\times 4$}\\
 &  PIP & Est. & S.D. &  PIP & Est. & S.D. &  PIP & Est. & S.D.\\
 \hline
 & \multicolumn{9}{c|}{Selection equation}\\
 \hline
(Intercept) & - & 0.813 & 0.020 & - & 0.813 & 0.020 & - & 0.813 & 0.020 \\

logc &  1.000 & -0.229 &  0.030 &  1.000 & -0.227 &  0.032 &  1.000 & -0.229 &  0.032 \\

idp &  0.094 & -0.001 &  0.012 &  0.115 & -0.001 &  0.013 &  0.082 & -0.004 &  0.012 \\

lpi & 0.919 & 0.072 & 0.028 & 0.897 & 0.069 & 0.029 & 0.870 & 0.071 & 0.030 \\

fmde & 0.090 & 0.001 & 0.023 & 0.118 & 0.000 & 0.027 & 0.111 & 0.002 & 0.027 \\

physlm & 0.998 & 0.099 & 0.024 & 0.981 & 0.098 & 0.027 & 0.988 & 0.099 & 0.025 \\

disea & 1.000 & 0.154 & 0.023 & 1.000 & 0.153 & 0.023 & 1.000 & 0.154 & 0.024 \\

hlthg & 0.020 & 0.000 & 0.004 & 0.016 & 0.000 & 0.003 & 0.019 & 0.000 & 0.007 \\

hlthf & 0.131 & 0.001 & 0.018 & 0.147 & 0.001 & 0.019 & 0.105 & 0.004 & 0.016 \\

hlthp & 0.844 & 0.077 & 0.038 & 0.916 & 0.081 & 0.033 & 0.809 & 0.075 & 0.038 \\

linc & 0.640 & 0.048 & 0.034 & 0.758 & 0.055 & 0.032 & 0.619 & 0.046 & 0.033 \\

lfam & 0.019 & 0.000 & 0.005 & 0.019 & 0.000 & 0.004 & 0.021 & 0.000 & 0.007 \\

educdec & 0.922 & 0.071 & 0.027 & 0.924 & 0.070 & 0.027 & 0.881 & 0.070 & 0.029 \\

xage &  0.031 & 0.000 &  0.007 &  0.033 & 0.000 &  0.006 &  0.028 & -0.001 &  0.008 \\

female & 1.000 & 0.196 & 0.024 & 1.000 & 0.195 & 0.025 & 1.000 & 0.196 & 0.024 \\

child & 0.057 & 0.001 & 0.011 & 0.054 & 0.000 & 0.011 & 0.046 & 0.002 & 0.011 \\

fchild &  1.000 & -0.143 &  0.024 &  1.000 & -0.142 &  0.024 &  1.000 & -0.144 &  0.024 \\

black &  1.000 & -0.227 &  0.021 &  1.000 & -0.225 &  0.021 &  1.000 & -0.227 &  0.021 \\
 \hline
 & \multicolumn{9}{c|}{Outcome equation}\\
 \hline
(Intercept) & - & 3.550 & 0.039 & - & 3.547 & 0.038 & - & 3.548 & 0.039 \\

logc &  1.000 & -0.229 &  0.027 &  1.000 & -0.230 &  0.027 &  0.998 & -0.229 &  0.029 \\

idp &  0.033 & -0.001 &  0.007 &  0.050 & 0.000 &  0.009 &  0.039 & -0.002 &  0.010 \\

lpi & 0.046 & 0.001 & 0.010 & 0.047 & 0.000 & 0.009 & 0.044 & 0.002 & 0.011 \\

fmde &  0.037 & 0.000 &  0.011 &  0.045 & 0.000 &  0.012 &  0.044 & 0.000 &  0.015 \\

physlm & 1.000 & 0.124 & 0.025 & 0.998 & 0.123 & 0.026 & 0.999 & 0.124 & 0.026 \\

disea & 1.000 & 0.205 & 0.027 & 1.000 & 0.205 & 0.027 & 1.000 & 0.206 & 0.027 \\

hlthg & 0.236 & 0.002 & 0.026 & 0.304 & 0.001 & 0.029 & 0.220 & 0.006 & 0.026 \\

hlthf & 0.740 & 0.066 & 0.040 & 0.823 & 0.071 & 0.038 & 0.714 & 0.066 & 0.040 \\

hlthp & 0.964 & 0.100 & 0.031 & 1.000 & 0.104 & 0.025 & 0.968 & 0.100 & 0.030 \\

linc & 0.997 & 0.138 & 0.034 & 1.000 & 0.145 & 0.031 & 0.999 & 0.137 & 0.033 \\

lfam &  0.853 & -0.079 &  0.037 &  0.924 & -0.082 &  0.032 &  0.777 & -0.075 &  0.040 \\

educdec & 0.076 & 0.001 & 0.014 & 0.088 & 0.000 & 0.015 & 0.070 & 0.003 & 0.014 \\

xage & 0.749 & 0.115 & 0.071 & 0.681 & 0.098 & 0.072 & 0.712 & 0.106 & 0.071 \\

female & 1.000 & 0.279 & 0.035 & 1.000 & 0.277 & 0.034 & 1.000 & 0.277 & 0.034 \\

child &  0.670 & -0.110 &  0.083 &  0.735 & -0.125 &  0.083 &  0.730 & -0.123 &  0.081 \\

fchild &  1.000 & -0.234 &  0.044 &  1.000 & -0.229 &  0.044 &  1.000 & -0.230 &  0.043 \\

black &  1.000 & -0.205 &  0.030 &  1.000 & -0.205 &  0.029 &  1.000 & -0.206 &  0.030 \\
 \hline
$\sigma$ & - & 1.571 & 0.029 & - & 1.573 & 0.029 & - & 1.572 & 0.030\\
 $\rho$ & - & 0.729 & 0.039 & - & 0.732 & 0.038 & - & 0.730 & 0.039 \\
\hline
\end{tabular}}\label{table:rand_spike}
\caption{Results from RAND data from varying the spike variances. The first column refers to what $\tau_{0,\alpha}$ and $\tau_{0,\beta}$ have been scaled by. The elicitation is otherwise the same as the previous data studies.}
\end{table}

\clearpage

\end{document}